\documentclass[proof]{pasj00}

\newcommand{\zph}  {\hbox{$z_{\rm ph}$}}

\newcommand{\sdrgs}{\hbox{sDRGs}}
\newcommand{\qdrgs}{\hbox{qDRGs}}
\newcommand{\gsim} {\lower.5ex\hbox{$\; \buildrel > \over \sim \;$}}
\newcommand{\lsim} {\lower.5ex\hbox{$\; \buildrel < \over \sim \;$}}
\newcommand{\psim} {\lower.5ex\hbox{$\; \buildrel \propto \over \sim \;$}}

\begin{document}
\SetRunningHead{Ma et al.}{Physical Properties of DRGs}
%\Received{}%{yyyy/mm/dd}
%\Accepted{}%{yyyy/mm/dd}
%\Published{}%{yyyy/mm/dd}

\title{Physical properties of distant red galaxies in the COSMOS/UltraVISTA field}

%%% begin:list of authors
% Do NOT capitalize all letters in "textsc".
\author{Zhongyang \textsc{Ma}}
\affil{Center for Astrophysics, University of Science and Technology of China, Hefei 230026, China; \\Key Laboratory for Research in Galaxies and Cosmology, CAS, Hefei 230026, China.}
\email{mzy@mail.ustc.edu.cn}

\author{Guanwen \textsc{Fang}
\thanks{Guanwen Fang and Zhongyang Ma contributed equally to this work.}}
\affil{Institute for Astronomy and History of Science and Technology, Dali University, Dali 671003, China; \\Key Laboratory for Research in Galaxies and Cosmology, CAS, Hefei 230026, China.}\email{wen@mail.ustc.edu.cn}

\author{Xu \textsc{Kong}}
\affil{Center for Astrophysics, University of Science and Technology of China, Hefei 230026, China; \\Key Laboratory for Research in Galaxies and Cosmology, CAS, Hefei 230026, China.}\email{xkong@ustc.edu.cn}

\and
\author{Lulu {\sc Fan}}
\affil{Center for Astrophysics, University of Science and Technology of China, Hefei 230026, China; \\Key Laboratory for Research in Galaxies and Cosmology, CAS, Hefei 230026, China.}
%%% end:list of authors

%%% Please use the following style in case that sorting by
%%% affiliation is impossible.
%
% \author{%
%   D-Firstname \textsc{D-Familyname}\altaffilmark{1}
%   E-Firstname \textsc{E-Familyname}\altaffilmark{1,2}
%   and
%   F-Firstname \textsc{F-Familyname}\altaffilmark{2}}
% \altaffiltext{1}{Address of Institute}
% \email{ddddd@xxx.xxx.xx.xx}
% \email{eeeee@xxx.xxx.xx.xx}
% \altaffiltext{2}{Address of Institute}

%% `\KeyWords{}' always has to be placed before `\maketitle'.
 %Do NOT move this preamble from here!

\maketitle

\begin{abstract}

We present a study on physical properties for a large
distant red galaxy (DRG) sample, using the $K$-selected multi-band
photometry catalog of the COSMOS/UltraVISTA field and the CANDELS NIR
data. Our sample includes 4485 DRGs with $(J-K)_\mathrm{AB}>1.16$ and
$K_\mathrm{AB}<$23.4 mag, and 132 DRGs have HST/WFC3 morphological measurements. The results of nonparametric measurements of DRG morphology are consistent with our rest-frame UVJ color classification: quiescent DRGs are generally compact while star-forming DRGs tend to have extended structures. We find
the star formation rate (SFR) and the stellar mass of star-forming DRGs
present tight ``main sequence'' relations in all redshift bins. Moreover,
the specific SFR (sSFR) of DRGs increase with redshift in all stellar mass
bins and DRGs with higher stellar masses generally have lower sSFRs, which
indicates that galaxies were much more active on average in the past, and
star formation contributes more to the mass growth of low-mass galaxies
than to high-mass galaxies. The infrared (IR) derived SFR dominate the
total SFR of DRGs which occupy the high-mass range, implying that the $J-K$
color criterion effectively selects massive and dusty galaxies. DRGs
with higher $M_{*}$ generally have redder $(U-V)_\mathrm{rest}$ colors,
and the $(U-V)_\mathrm{rest}$ colors of DRGs become bluer at higher
redshifts, suggesting high-mass galaxies have higher internal dust
extinctions or older stellar ages and they evolve with time. Finally,
we find that DRGs have different overlaps with EROs, BzKs, IEROs and high-$z$
ULIRGs indicating DRGs is not a special population and they can also be
selected by other color criteria.

\end{abstract}

\section{Introduction}

The universe at high redshift looks very different from that at low
redshift and local.
At the epoch of $z\sim2$, a large fraction of massive galaxies are still
active, and their morphology are quite different from that of nearby galaxies
\citep{McCracken2010, Fang2012, Wang2012}. In
order to examine current theoretical models of galaxy formation and
evolution, studies of galaxies over a wide range of lookback time are
needed \citep{Shapley2011}. Within the past decade, many novel techniques
have been applied to assemble multi-wavelength observations of galaxies
at this important epoch and significant progress has been made in our
understanding of high-redshift galaxies \citep{Steidel1996, Smail1997,
Franx2003, Daddi2004, Gronwall2007, Yan2007, Dey2008,  Dunne2009,
Caputi2012, Arcila2013, Ilbert2013}.

At early time, one method for selecting high-redshift galaxies is the
Lyman break technique, which uses the significant break at 912 \AA\,
in the galaxy rest-frame UV spectrum to select star-forming galaxies
(so-called Lyman Break Galaxies, or LBGs), and was first applied at
$z\sim3$ \citep{Steidel1996}. The LBGs are UV-bright star-forming galaxies
and dominated by relatively low-mass systems ($\sim10^{10}M_{\odot}$). Consider that
most of local elliptical and S0 galaxies commonly have stellar masses great than $10^{11}M_{\odot}$, the LBGs may not be the progenitors of these massive
galaxies today \citep{Conselice2007}, and the Lyman break technique have missed
a significant population of red massive galaxies at high redshift, such as
distant red galaxies (DRGs) \citep{Franx2003}, BzKs \citep{Daddi2004},
ultraluminous infrared galaxies (ULIRGs) \citep{2003Natur.422..695C}, etc.
As a
supplement to these UV-bright LBGs, \cite{Franx2003} introduced a new
photometric technique for selecting $z>2$ massive red galaxies. The
relevant spectral break is the Balmer break at 3650 \AA{} which attributes
to the spectral shape of young stars and the 4000 \AA{} break which
is caused by the absorption of metals in old stars. Since these breaks
shift into the $J$ band at $z\sim2$, a red $J-K$ color is a simple and
effective criterion \citep{Franx2003}. They simulated the tracks in the
$J-K$ color versus redshift diagram with model spectra taken from Bruzual
\& Charlot 1993, and imposed a criterion $(J-K)_\mathrm{Vega}>2.3$
for selecting galaxies at $z>2$ (so-called Distant Red Galaxies, or DRGs).

DRGs commonly have higher star formation rates (SFRs)
($\sim100M_{\odot}/$yr) and more dust extinctions ($A_\mathrm{V}=1-3$
mag) \citep{Franx2003, FS2004, Papovich2005, Webb2006}. The redshift
distribution of DRGs is quite broad, and the $(J-K)_\mathrm{Vega}>2.3$
selection criterion does not uniquely sample high-mass galaxies
\citep{Papovich2005, Grazian2006, Conselice2007}. However, DRG selection
technique does locate very massive galaxies ($\sim10^{11-12}M_{\odot}$)
at $z>2$ \citep{vanDokkum2006, Conselice2007}. In the present-day
Universe, most of massive ($M_{*}>10^{11}M_{\odot}$) galaxies are
quiescent systems. On the contrary, at $z\sim2$ a large proportion of
massive galaxies are still active, while others are ``red and dead''
with little or no ongoing star-forming activities \citep{Labbe2005,
Papovich2006, Daddi2007, Wang2012}. Transformation from the active
systems into the evolved systems is a reasonable explanation to this
dramatic evolution. Some studies suggested that DRGs show properties of
both passive systems and dust obscured active systems \citep{FS2004,
Labbe2005, Papovich2005, Webb2006}. \cite{Labbe2005} showed color
tracks for stellar population models \citep{BC03} in the $(I-K)$
vs. $(K-[4.5])$ diagram, and found that 70\% of DRGs are well fit with
dusty star-forming models and 30\% are best described by old stellar
population models. \cite{Webb2006} combined Spitzer 24$\mu$m flux, and
found a similar percentage ($\sim$65\%) of DRGs are dusty active galaxies. By studying the spectra, \cite{Kriek2006a, Kriek2006b} found that the stellar populations differ greatly among DRGs, from dusty starburst with small break to evolved galaxy with strong break and no line emission, and both the H$\alpha$ measurements and stellar continuum modeling imply 45\% of $K$-selected massive galaxies at $z\sim2$ are not forming stars intensely.

The morphologies of DRGs are diverse. \cite{Toft2005} analyzed the
morphologies of DRGs in the Hubble Ultra Deep Field based on HST
Advanced Camera for Surveys (ACS) images and Near Infrared Camera and
Multi-Object Spectrometer (NICMOS) images, and found that the rest-frame
optical morphology of DRG is quite different from the rest-frame UV
morphology. In the near-infrared images, some DRGs are extended and
others are compact. \cite{Conselice2007} examined the morphologies of
DRGs in the Extended Groth Strip (EGS) field based on ACS F814W images
with eye-ball and nonparametric classification method, and found that
the morphologies of DRGs show a diversity of types at all redshifts,
from elliptical, peculiar to disk.

The $(J-K)_\mathrm{Vega}>2.3$ criteria is a simple but outstanding selection method utilising the 3650 and 4000 \AA breaks for high-$z$ massive red galaxies, which are supplements to those UV-bright low-mass LBGs. The DRG sample provides us with the chance to investigate the nature of the population of the massive red objects at high redshifts, which are supposed to be the progenitors of local massive populations, thus it makes us better understand the evolution of massive galaxies. Since the DRG samples of previous research programs are commonly small ($<1000$), it still remains many uncertainties on the physical properties of distant red galaxies. In order to have a thorough understanding of this kind of galaxies, in this
paper we present the largest sample of DRGs so far, which is selected from
the COSMOS/UltraVISTA field, based on the high quality NIR data presented
by \cite{McCracken2012} and \cite{Muzzin2013a}. To clearly describe the details of the characteristics of DRGs, we investigate the correlations between their SFRs, redshifts, masses and morphologies. The rest-frame UV emission of galaxies is mainly contributed by the hottest stars and can be severely affected by dust extinction, therefore it is essential to study $z\sim2$ galaxies from observed NIR bands, which probe the rest-frame optical morphologies. Our morphology analysis was enabled by the HST Wide Field Camera 3 (WFC3) NIR imaging of the Cosmic Assembly Near-infrared Deep Extragalactic Legacy Survey (CANDELS)\citep{Grogin2011, Koekemoer2011}, and we compare the nonparametric measure of galaxy morphology with visual inspection and UVJ color-color diagram. In the meanwhile, we also compare DRG with other color selection criteria.

The paper is organized as follows. We introduce the $K$-selected catalog of the COSMOS/UltraVISTA field in Section 2, and the DRG sample selection method and classification in Section 3. We show our results on physical properties of DRGs in
Section 4. We compare the DRG method with  other color selection
methods in Section 5, and conclude our results in Section 6. Throughout
this paper, we assume an $\Omega_\mathrm{M}=0.3, \Omega_{\Lambda}=0.7$
and $H_{0}=70$ km s$^{-1}$ Mpc$^{-1}$ cosmology, and all magnitudes and
colors are given in AB system unless stated otherwise. Symbol such as
``[3.6]'' means the AB magnitude at the wavelength 3.6 $\mu$m.

%\newpage

\section{Observations and data}% section2

The COSMOS field is centered at $\alpha = 10^{h}00^{m}28.6^{s}$,
$\delta = +02^{d}12^{m}21^{s}$ (J2000.0), and covers an area of
$\sim2$ square degree \citep{Scoville2007}.
The COSMOS project gathers a variety of ground- and space-based
observations, covering nearly the whole spectrum range
\citep{Martin2005,  Capak2007, Hasinger2007, Sanders2007, Schinnerer2007,
Taniguchi2007, Lilly2009, Ilbert2010, McCracken2010}.

The multi-band photometry data we used in this paper come from the $K$-selected catalog of the COSMOS/UltraVISTA field provided by \cite{Muzzin2013a}, which is
produced based on the near infrared (NIR) data from the UltraVISTA data
release 1 \citep{McCracken2012}.
The high-quality and deep-enough NIR data make
it possible for us to select a large sample of DRGs. The
catalog contains PSF-matched photometry in 30 photometric bands
covering the wavelength from 0.15$\mu$m to 24$\mu$m, and includes the
available GALEX \citep{Martin2005}, CFHT/Subaru \citep{Capak2007},
UltraVISTA \citep{McCracken2012}, S-COSMOS \citep{Sanders2007}, and
zCOSMOS \citep{Lilly2009} datasets. After removing regions contaminated
by bright stars, the effective area of overlap between optical and NIR
data is 1.62 deg$^{2}$. The photometry in all bands is corrected for
Galactic dust attenuation, and stars are separated from galaxies using
the $(J-K)$ vs. $(u-J)$ color space, and it agrees well with
{\tt SExtractor}'s {\tt class$\_$star} parameters.
The catalog is presented as a set of fluxes
in the $2\farcs1$ color aperture with an AB zeropoint  of 25.0, and the
90$\%$ completeness limit of the catalog is $K_\mathrm{s,tot}$ = 23.4.

The derived physical parameters we used in this paper also come from the catalog provided by \cite{Muzzin2013a}. Photometric redshifts are derived for all galaxies using the
{\tt EAZY} code \citep{Brammer2008} with 30 bands, and have been
calibrated by the highest-quality spectroscopic redshifts from
zCOSMOS. The stellar population parameters such as stellar mass
($M_{*}$), age, $A_\mathrm{V}$, SFR$_\mathrm{SED}$ are derived by SED fitting method
in {\tt FAST} code \citep{Kriek2009a}. The rest-frame UV luminosity is derived by {\tt EAZY}. {\bf The total IR luminosity is derived by fitting the 24$\mu$m flux using the log-average of the \cite{DH02} templates, but note that using the log-average of the \cite{CE01} templates would provide very similar results.} Then the rest-frame UV and the total IR luminosities are converted to SFR$_\mathrm{UV,uncorr}$ and SFR$_\mathrm{IR}$ according to \cite{Kennicutt1998}. The total star formation rate of the galaxy can then be determined via SFR$_\mathrm{tot} =$ SFR$_\mathrm{UV,uncorr} +$ SFR$_\mathrm{IR}$. The rest-frame $U-V$ and $V-J$ colors are calculated for all galaxies by {\tt EAZY}, which determines the colors by integrating the best-fit SED through the redshifted filter curves over the appropriate wave length range.

\section{Selection and classification of DRGs}%section3

\subsection{DRG Sample Selection}%3.1

To select a sample of DRGs from UltraVISTA catalog, firstly we remove all
the objects  classified as stars and contaminated galaxies defined by
\cite{Muzzin2013a}. In order to make our photometric selecting criterion close to the original DRG selecting criterion introduced by \cite{Franx2003}, we make a simulation in BC03 code \citep{BC03} to determine the corrections for $J$ and $K$ bands. Finally, we set a small correction for $J-K$ color: $(J-K)_\mathrm{UltraVISTA}+0.14=(J-K)_\mathrm{ISAAC}$ (AB), so we set $(J-K)_\mathrm{UltraVISTA}>1.16$ (AB), which is equivalent to $(J-K)_\mathrm{ISAAC}>1.3$ (AB), as the criterion for selecting DRGs. On the other hand, we carefully compared the location of the stellar sequence with previous works, and find the distance on $J-K$ color between stellar sequence and DRG selecting criterion is 1.36, which is consistent with previous works \citep{Foucaud2007, Kajisawa2008}.

Figure 1 shows the $K$-selected galaxies (objects in masked area
are removed) in the $(J-K)$ vs. $K$ diagram, the dashed line represents
$J-K=1.16$, which is our DRG selecting criterion, objects above which are
all DRGs and are plotted in red color. The dot-dashed line represents the stellar sequence, and all stars are plotted in green color. The total number of DRGs we select from the COSMOS/UltraVISTA field is 4485 with $K<23.4$ and $J<24.2$ mag {\bf (Therefore, all DRGs in our sample actually have $K<24.2-1.16=23.04$ mag).}

\begin{figure}%Fig.1
\includegraphics[angle=-90,width=0.95\columnwidth]{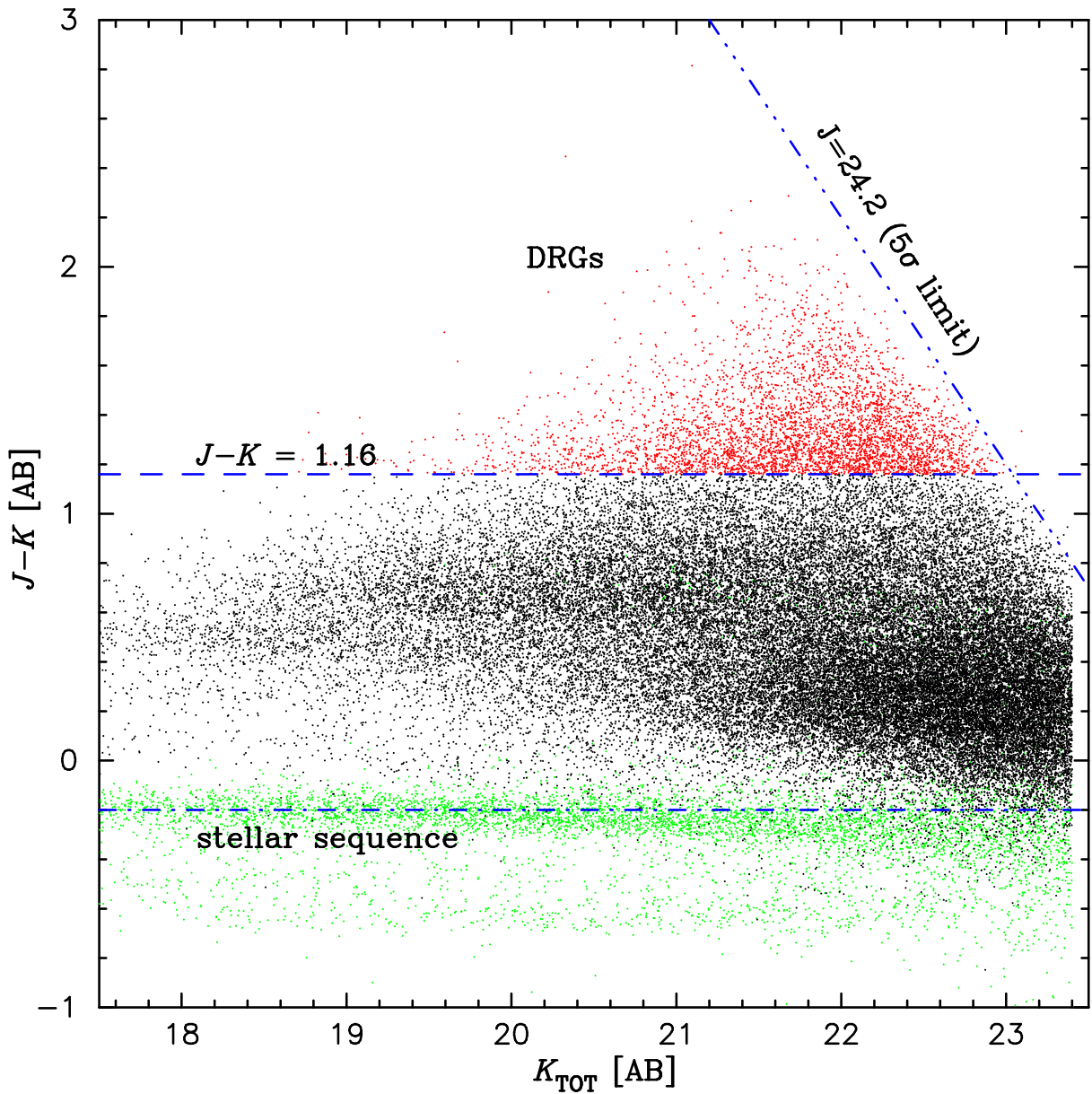}
\caption{$(J-K)$ color vs. $K$ magnitude for galaxies in the
COSMOS/UltraVISTA field.
The dashed line corresponds to $J-K=1.16$ which is our DRG selecting
criterion, galaxies above which are all DRGs and are plotted in red color.
The dot-dashed line represents the stellar sequence, and stars are
plotted in green. The dot-dot-dashed line indicates the depth of the
$J$ band.}
\end{figure}

\begin{figure}%Fig.2
\includegraphics[angle=-90,width=0.95\columnwidth]{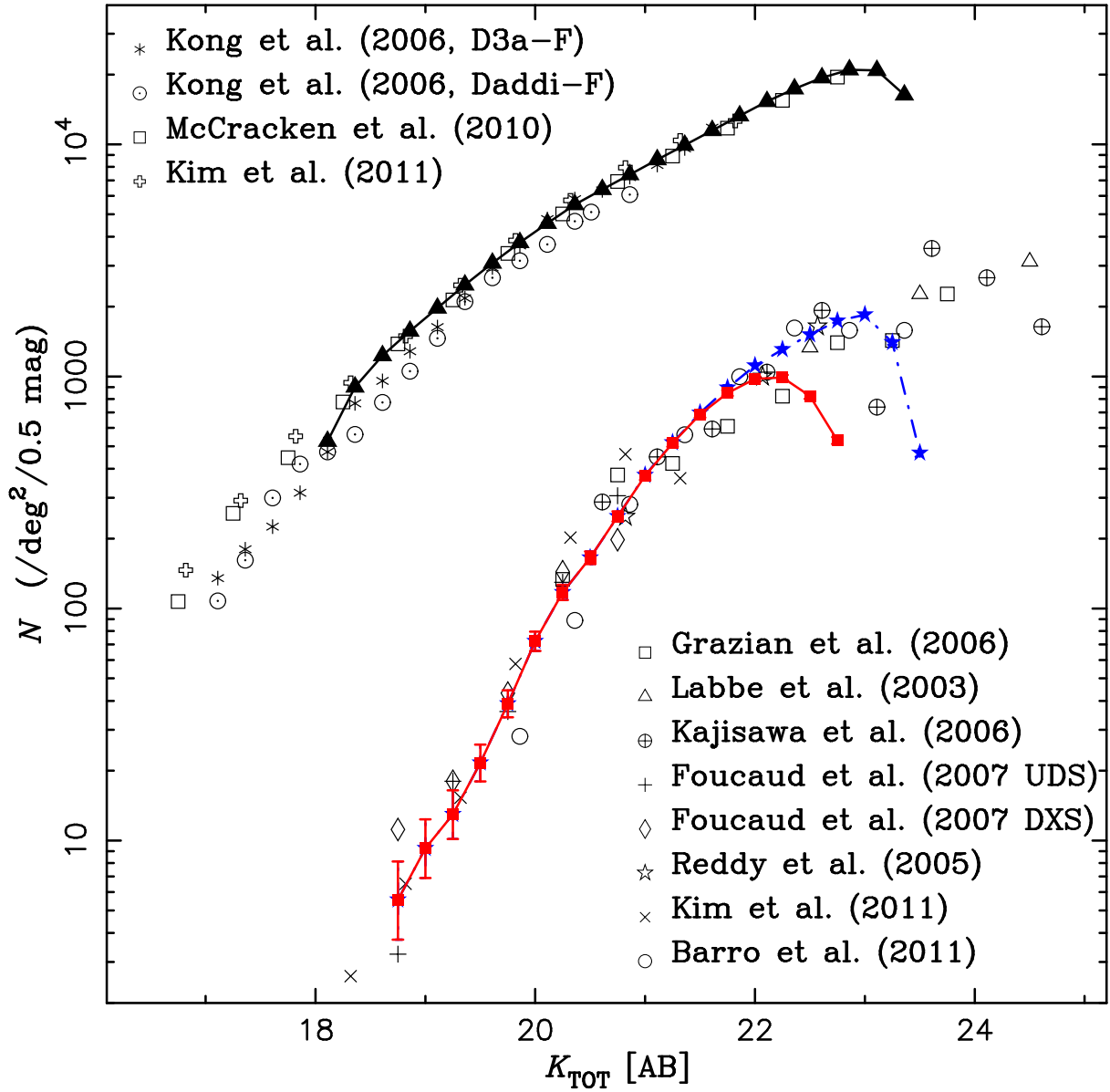}
\caption{$K$-band differential number counts for DRGs (red squares)
and all $K$-selected galaxies (black triangles) in the COSMOS/UltraVISTA field.
The blue stars represent the number counts of DRGs regardless of their 5$\sigma$ detection in $J$-band or lack there of. For comparison, we also overplot
the number counts of DRGs and all $K$-selected galaxies from previous works.}
\end{figure}

Figure 2 shows the $K$-band differential number counts of DRGs and  all $K$-selected galaxies. For comparison, we have also plotted the number count results
from other literature. In this figure, the red squares and the black
triangles represent the number counts of DRGs and all $K$-selected galaxies in the
COSMOS/UltraVISTA field, respectively. The blue stars represent the number
counts of DRGs whose $J$-band magnitude limit is not considered. Results
of DRGs from GOODS-S \citep{Grazian2006}, HDF-S \citep{Labbe2003},
GOODS-N \citep{Kajisawa2006}, UDS-EDR \& DXS-EDR \citep{Foucaud2007},
GOODS-N \citep{Reddy2005}, UKIDSS \citep{Kim2011} and EGS \citep{Barro2011}
are also plotted. And results of all $K$-selected galaxies from Deep3a-F \& Daddi-F
\citep{Kong2006}, COSMOS \citep{McCracken2010} and UKIDSS \citep{Kim2011}
are plotted in different symbols as well. As shown by this figure, our
number counts of DRGs and all $K$-selected galaxies in the COSMOS/UltraVISTA field
are in good agreement with those in previous works in other fields.

\subsection{Classification based on rest-frame UVJ colors}%3.2

As shown by previous literatures \citep{Labbe2005, Webb2006, Kriek2006a, Kriek2006b}, DRG is a
diverse population, most of the DRGs are dusty star-forming galaxies
while others are old and red with little on going star formation
activities.
In order to specify the details of the physical properties and the
evolution of them, separating DRGs into star-forming and quiescent
populations is necessary.

We classify our sample as star-forming DRGs (sDRGs) and quiescent DRGs
(qDRGs) based on the rest-frame UVJ color criteria defined by
\cite{Muzzin2013b}.
According to their definition, qDRGs are classified by:
\begin{equation}
U-V>1.3,\ V-J<1.5,\ [\mathrm{all\ redshifts}],
\end{equation}
\begin{equation}
U-V>(V-J)\times0.88+0.69,\ [0.0<z<1.0],
\end{equation}
\begin{equation}
U-V>(V-J)\times0.88+0.59,\ [1.0<z<4.0].
\end{equation}
Otherwise, they are star-forming DRGs. These classification criteria are
plotted as dashed lines in Figure 3. Based on this classification method,
we identify 3725 sDRGs and 760 qDRGs, the percentage of sDRGs is 83\%,
which is a little higher than previous results \citep{Labbe2005, Webb2006}.

\begin{figure*}% Fig.3
\includegraphics[angle=-90,width=0.95\textwidth]{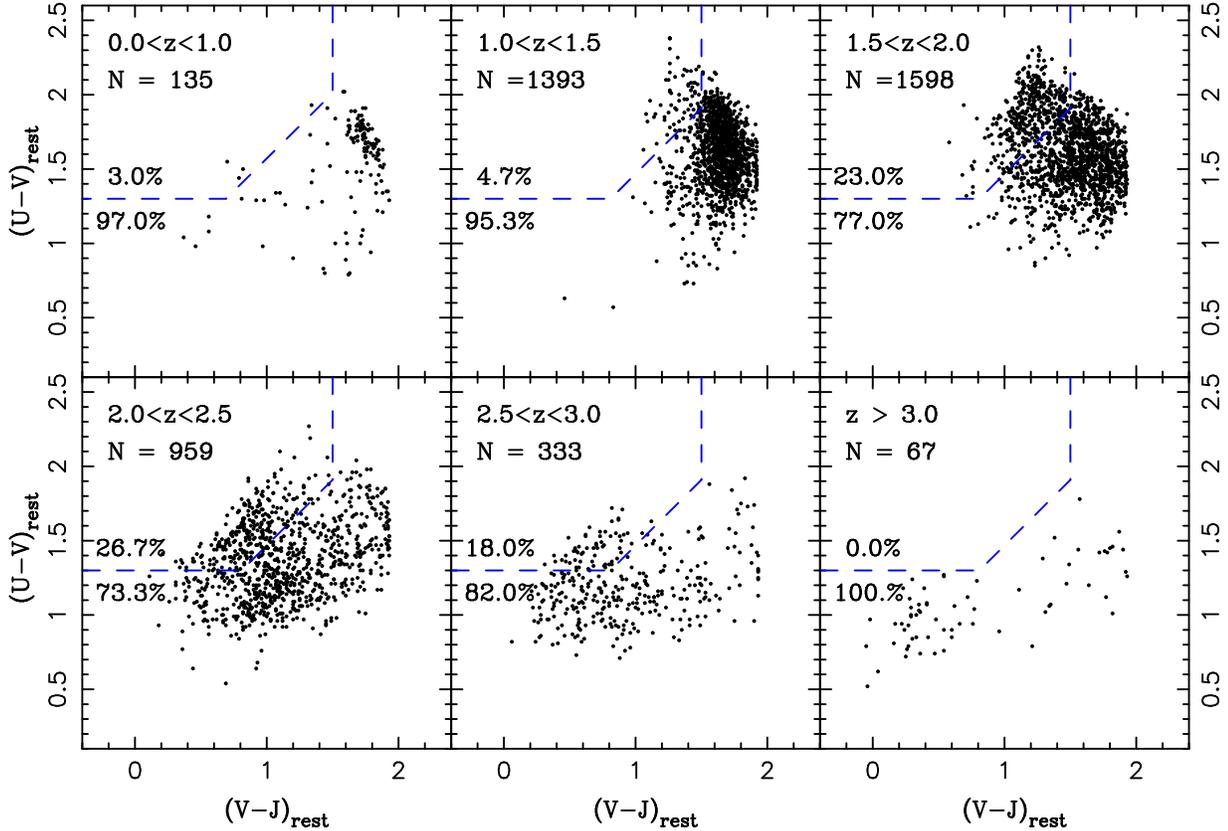}
\caption{Classification for DRGs based on rest-frame $(U-V)$ vs.
$(V-J)$ colors in six redshift bins.
This color criteria (dashed lines) used to separate star forming DRGs
(sDRGs) and quiescent DRGs (qDRGs) are defined by Muzzin et al. 2013b.
The number of DRGs, the fraction of sDRGs and qDRGs in each redshift
bins are shown, respectively.}
\end{figure*}

\subsection{Classification based on visual morphologies}%3.3

Morphologies of DRGs provide direct information on the formation and
evolution history of these objects, and correlate a range of physical
properties such as stellar mass, SFR and rest-frame color.
It is difficult to study morphologies of high redshift dusty galaxies
based on their observed optical images, because their observed optical
light probes the rest-frame UV emission for objects at $z\sim2$, and
their apparent morphologies can be easily changed by patchy dust
extinction. For example, \cite{Conselice2007} showed that DRGs have peculiar, clumpy
morphologies in the $HST$/ACS F814W band image.
Based on the $HST$ NICMOS-$JH$ ($0\farcs09$/pixel) and ACS-$BViz$
($0\farcs03$/pixel) images, however, \cite{Toft2005} found that the
rest-frame optical morphology of DRGs is quite different from the
rest-frame UV morphology.
The rest-frame UV emission of galaxies mainly contributed by the hottest
stars and can be severely affected by dust extinction, therefore it is
essential to study $z\sim2$ galaxies from observed NIR bands which probe
the rest-frame optical morphologies.
Our morphology analysis was enabled by the $HST$/WFC3 NIR imaging with a
resolution of $0\farcs06$/pixel from the Cosmic Assembly Near-infrared
Deep Extragalactic Legacy Survey (CANDELS)
\citep{Grogin2011, Koekemoer2011}.

Since the redshift distribution of DRGs is quite broad, we analyze their
rest-frame optical morphologies on WFC3 F125W ($J$) or F160W ($H$) bands
according to their redshifts.
For DRGs with $1<z<1.6$, we choose WFC3 F125W ($J$) bandpass for
morphological analysis, it corresponds approximately to $V$-band in the rest-frame in
this redshift range, but in the redshift range of $1.6<z<3$, we analyze
galaxy morphology in the rest-frame $V$ from the F160W ($H$) image instead.
Finally, 37 DRGs in our sample have $J$-band counterparts ($1<z<1.6$), and
95 DRGs ($1.6<z<3$) are detected at $H$-band image. The sensitivity (5$\sigma$) of the CANDELS imaging data reach 27 mag in the F125W and the F160W, and all of DRGs in our sample are brighter than 24 mag in both $J$ and $H$ bands, thus the CANDELS data is deep enough to morphologically classify this DRG sample.

We perform the visual inspection for this sample by three of us independently, we combined each classifier's results and review the images together to resolve the disagreements. The disagreements cover about 10\% of this sample, and mainly exist among galaxies which can't be clearly classified as disks or irregular galaxies. We find that
the DRGs in our sample show very diversified morphologies, covering
a wide range of types from extended disks, compact spheroids to clumpy
and irregular morphologies. We define three general morphological types
as follows:
\begin{description}
\item[Spheroid]: round and centrally concentrated source with no extended
structure.
\item[Disk]: single bright core with smooth extended disk-like structure.
\item[Irregular]: clumpy structure with more than two bright cores,
sometimes shows interaction features such as tidal arms.
\end{description}

Examples of three morphological types are shown in Figure 4. The result of the visual inspection supports the finding that the significant evolution of the Hubble sequence has already existed at high redshift \citep{Kajisawa2001}. In Table
1 we list the numbers of star-forming and quiescent DRGs classified
by rest-frame UVJ colors in three visual morphological types. Among
star-forming DRGs, the spheroids occupy a proportion of 19.8\%, disks occupy 32.7\%
and irregulars make up 47.5\%. And among quiescent DRGs, the spheroids
make up 77.4\%, disks make up 16.1\% and irregulars make up 6.5\%. We
find that most of quiescent DRGs are compact and star-forming DRGs tend
to have extended structures.

\begin{table*}
\begin{center}
\caption{Numbers of star-forming and quiescent DRGs, classified
by rest-frame UVJ colors in three visual morphological types.}
\begin{tabular}{lrrrr}
\hline
DRGs ($1<z<3$)  & Spheroid & Disk & Irregular & Total\\
\hline
Star-forming DRGs (\sdrgs) & 20 & 33 & 48 & 101 \\
Quiescent DRGs (\qdrgs) & 24 & 5 & 2 & 31 \\
\hline
\end{tabular}
\end{center}
\end{table*}

\begin{figure*}% Fig.4
\centering
\includegraphics[width=0.9\textwidth]{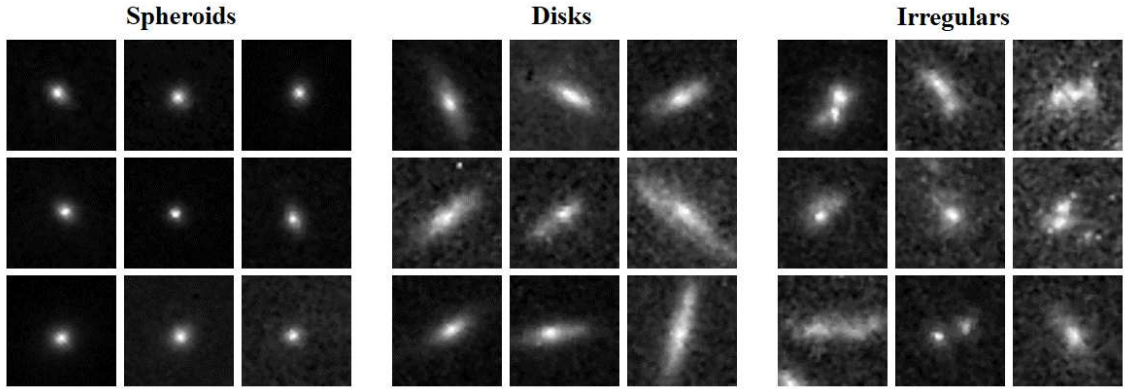}
\caption{Examples of three general morphological types from visual
inspection: spheroid on the left, disk in the middle and irregular on
the right panel. The size of each image is $2\farcs4\times2\farcs4$.}
\end{figure*}

\section{Results}%section4

\subsection{Redshift and mass distributions}%4.1

The photometric redshifts from the UltraVISTA catalog are in good agreement
with the spectroscopic redshifts from zCOSMOS \citep{Lilly2009}, with an
average $\delta z/(1+z)=0.013$ \citep{Muzzin2013a}. We show the redshift
distributions of both star-forming and quiescent DRGs in the left panel of
Figure 5. The star-forming DRGs (blue lines) distribute at $0<z<3.5$ with
$\langle z \rangle=1.74$, and the quiescent DRGs (red lines) are in the
range $z\sim1-3$ with $\langle z \rangle=1.98$. For all DRGs, our sample
has $\langle z \rangle$=1.79. In total, there are 3\% DRGs distribute at
$z<1$, 95.5\% at $1<z<3$ and 1.5\% at $z>3$. The right panel of Figure 5
shows the stellar mass distributions of both star-forming and quiescent
DRGs. We find the masses of them are relatively high, and mainly range
from $10^{10}M_{\odot}$ to $10^{11.5}M_{\odot}$. The mean stellar masses
are $10^{10.7}M_{\odot}$, $10^{10.6}M_{\odot}$ and $10^{10.8}M_{\odot}$
for all, star-forming and quiescent DRGs in our sample, respectively.

\begin{figure*}% Fig.5
\includegraphics[angle=-90,width=0.95\textwidth]{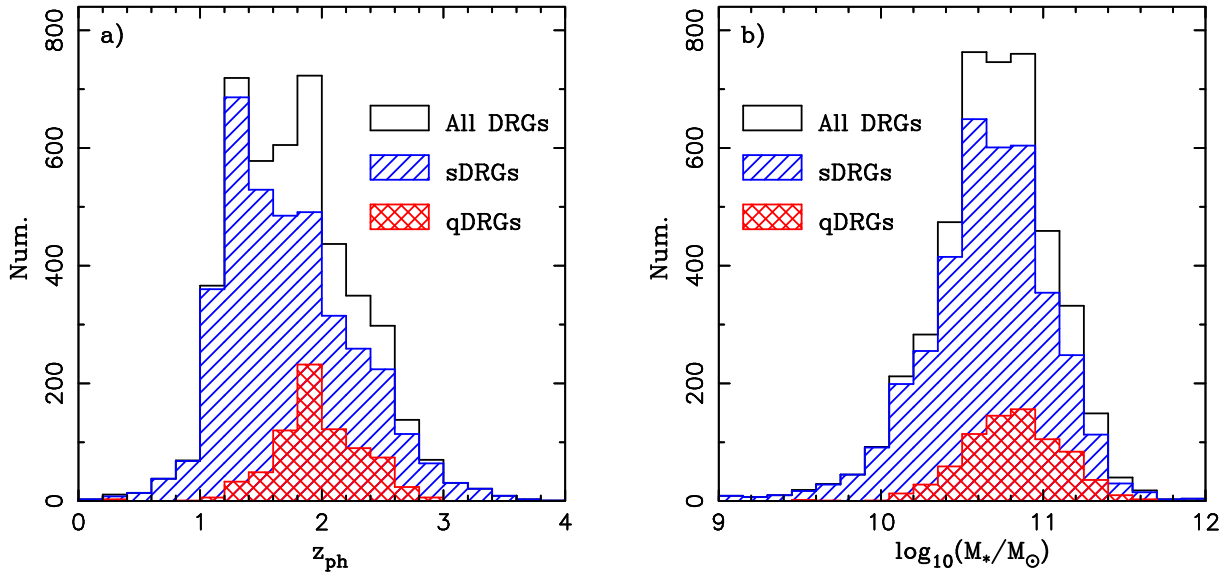}
\caption{Photometric redshift distribution (left panel) and stellar mass
distribution (right panel) of DRGs. All DRGs are plotted in black lines,
while star-forming and quiescent DRGs classified by rest-frame UVJ color
are plotted in blue and red, respectively.}
\end{figure*}

\begin{figure}% Fig.6
\includegraphics[angle=-90,width=0.95\columnwidth]{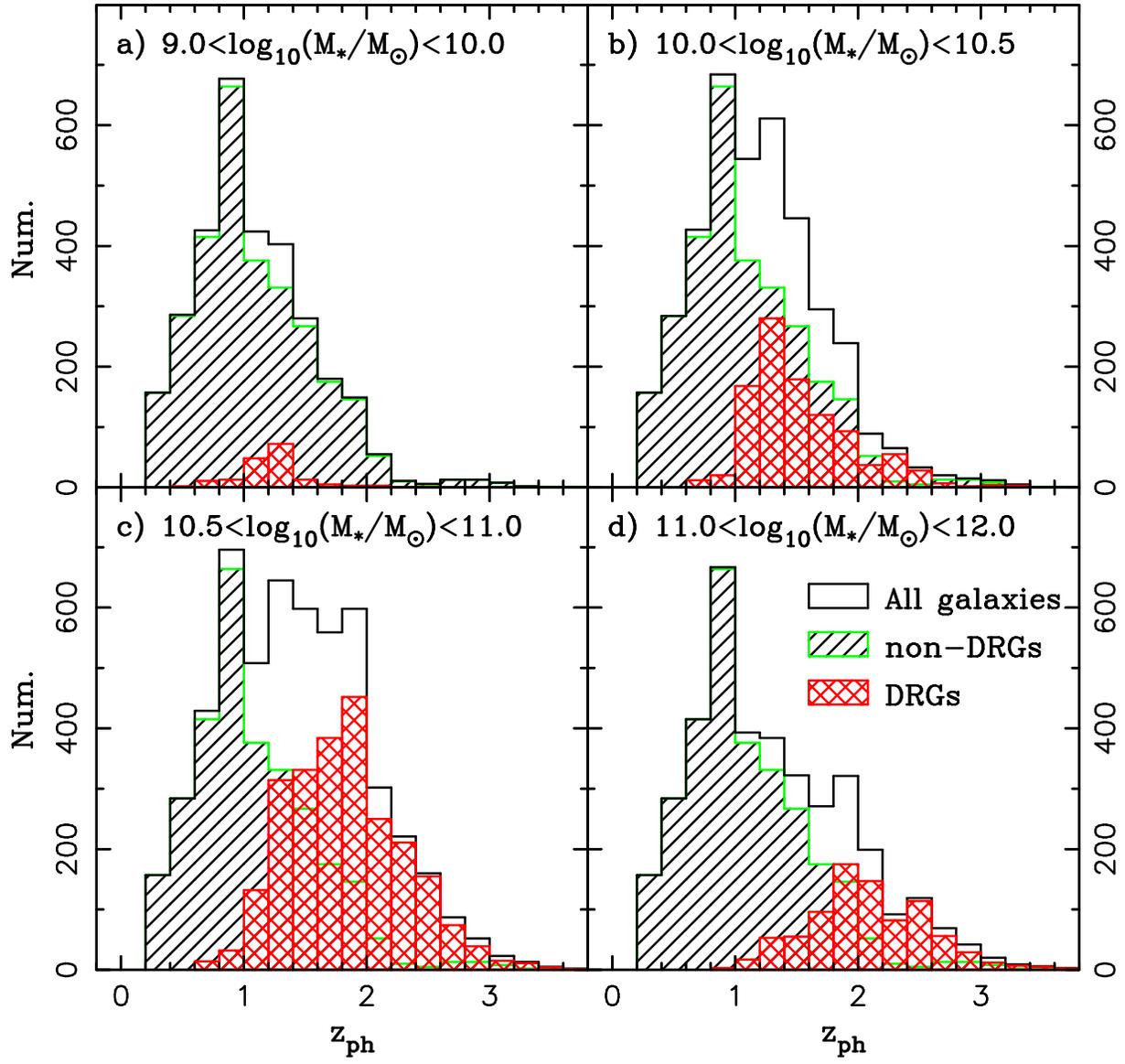}
\caption{Redshift distributions of all galaxies (black lines), DRGs (red
lines) and {\bf non-DRGs (green lines)} in the COSMOS/UltraVISTA field in four stellar mass
bins.}
\end{figure}

We compare the redshift distributions of DRGs with those of all of
galaxies
in four stellar mass bins in Figure 6. We find that in higher mass bins
(e.g., $10^{10.5}M_{\odot}-10^{11}M_{\odot}$,
$10^{11}M_{\odot}-10^{12}M_{\odot}$), DRGs represent nearly all of
galaxies at $z>2$.
At $M_{*}>10^{10.5}M_{\odot}$ and $z>2$, the fraction of DRGs out of all
galaxies is 70\%, which is in good agreement with \cite{vanDokkum2006}.
From the analysis above, we demonstrate that although DRG selecting
criterion also sample low-mass galaxies at low redshifts, DRGs occupy
large proportion of galaxies at high-mass and high-$z$ range, implying
that DRG selecting criterion is sensitive to massive galaxies at high
redshifts. Our results are also demonstrated in \cite{Conselice2007}. The spectroscopy study of \cite{Kriek2008} also supports our finding that red galaxies dominate the high-mass end of the galaxy population at $z\sim2-3$.

We have checked the mass completeness for our DRG sample, and find nearly all DRGs are above $10^{10}M_{\odot}$ and the majority of them are above $10^{10.5}M_{\odot}$ at $1<z<3$. The distance between the majority of DRGs and the 90\% completeness boundary at $1<z<3$ are safe enough, which ensures our DRG is a complete sample. Only a very small proportion of DRGs are close to the completeness boundary, and most of them are located in low-mass and low-$z$ range.

\subsection{Nonparametric morphology}%4.2

To further describe the morphologies of these sources, we have performed
nonparametric measures of galaxy morphology, such as Gini coefficient
(the relative distribution of the galaxy pixel flux values, or Gini) and
$M_{20}$ (the second-order moment of the brightest 20\% of the galaxy's
flux) \citep{Lotz2006}, using the Morpheus-software developed by Bob
Abraham \citep{Abraham2003}. Figure 7 shows the distribution of DRGs on
the Gini$-M_{20}$ plane in F125W and F160W bands, which corresponds to the
redshift range $1<z<1.6$ and $1.6<z<3$, respectively. Diamonds, stars
and triangles represent spheroid, disk and irregular galaxies, classified
by visual inspection. Star-forming and quiescent DRGs classified by
the rest-frame UVJ colors are plotted in blue and red, respectively. We
find that most of quiescent DRGs are visually compact, and have larger
Gini and smaller $M_{20}$ values, while the star-forming DRGs contain
different kinds of morphological types but most of them are visually
extended and have smaller Gini and larger $M_{20}$ values.
Similar results are also reported in \cite{Lee2013} and
\cite{Bassett2013}.

\begin{figure*}% Fig.7
\centering
\includegraphics[angle=-90,width=0.95\textwidth]{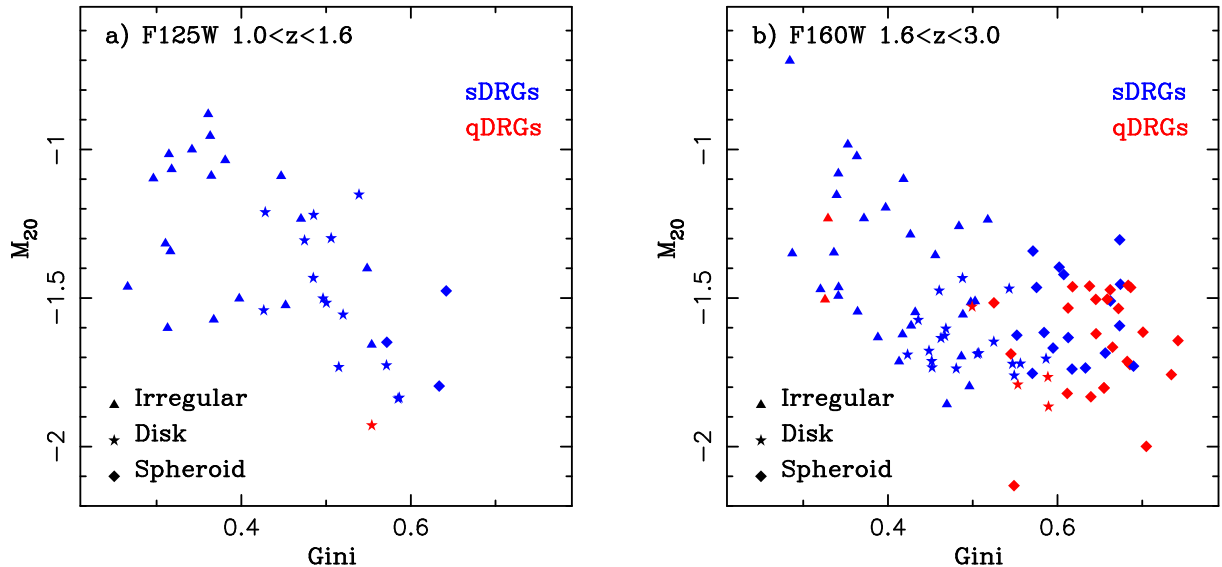}
\caption{ {\bf $M_{20}$ vs. Gini coefficient for DRGs}. The star-forming DRGs (sDRGs)
are plotted in blue color, while quiescent DRGs are in red.
Spheroid, disk and irregular DRGs which are classified by visual
inspection, are plotted in diamond, star and triangle symbols,
respectively.
DRGs on left panel with redshift $1<z<1.6$, whose morphological
parameters are measured from F125W image. DRGs on right panel distribute
 at $1.6<z<3$, whose morphological parameters are measured from F160W
image.}
\end{figure*}

\cite{Kriek2009a, Kriek2009b} support our morphological results that quiescent galaxies are commonly found to be ultra-compact, while galaxies with high SFRs have irregular and clumpy morphologies. Our morphological results also support the evolutionary tracks described in
\cite{Barro2013}, \cite{Bedregal2013} and \cite{Tadaki2014} that compact
DRGs (with larger Gini and smaller $M_{20}$) contain compact quiescent
galaxies (cQGs) and their progenitors compact star-forming galaxies
(cSFGs). In our sample, there are 54.5\% (24/44) spheroidal DRGs are quenched, namely they are cQGs, and the rest of them are cSFGs. The cSFGs are formed by gas-rich processes such as major mergers
at high redshifts, and rapidly quenched (probably by AGN) into cQGs,
most of which are classified as spheroidal quiescent DRGs. These compact QGs at $z\sim2$ may be the cores of today's massive elliptical galaxies, as described by \cite{vdSande2013} and \cite{Patel2013}, these quiescent galaxies at $z\sim2$ grow inside-out, consistent with the expectations from minor mergers.

\begin{figure*}% Fig.8
\includegraphics[angle=-90,width=0.95\textwidth]{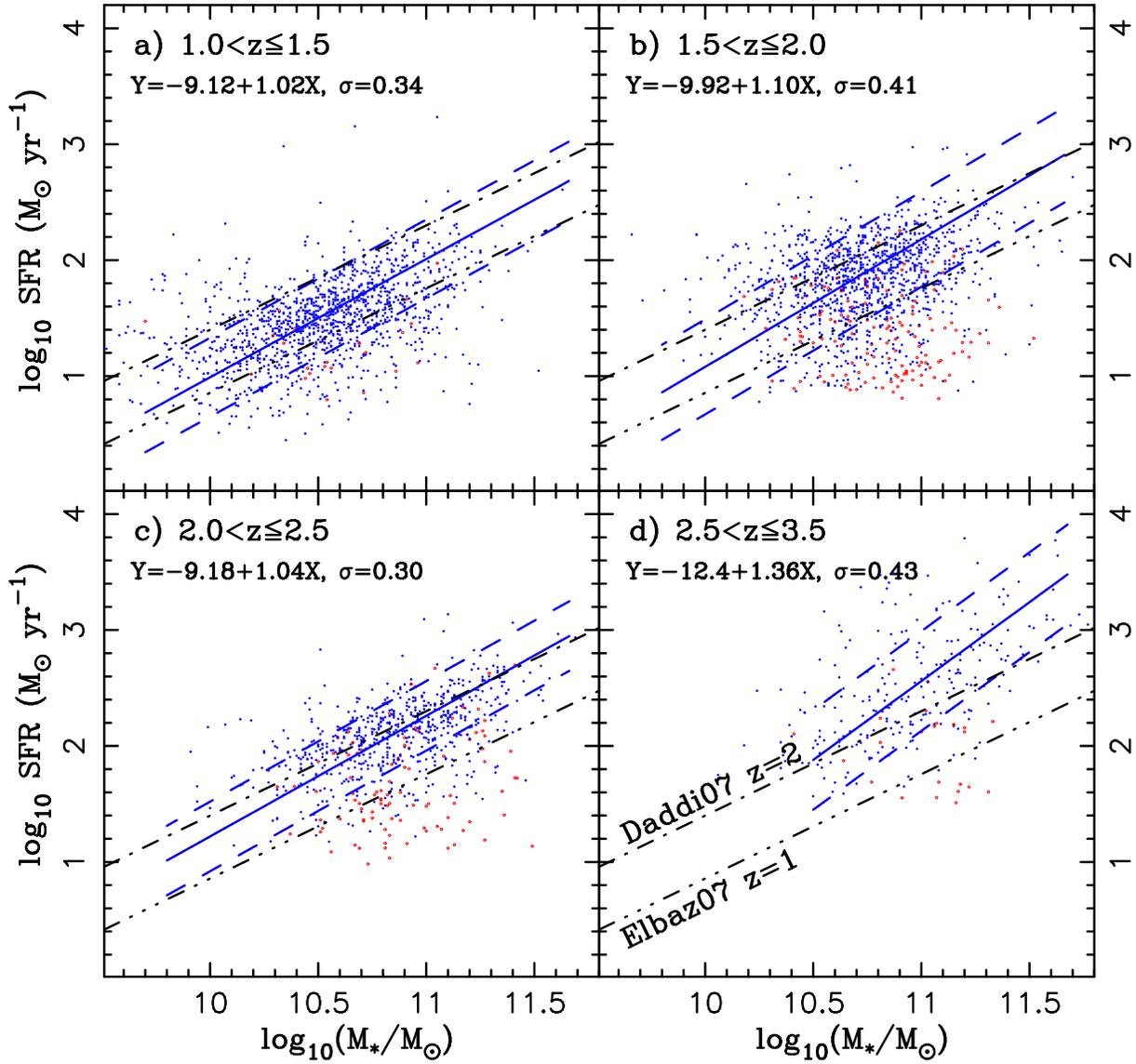}
\caption{ {\bf SFR vs. $M_*$ diagram for DRGs} in four redshift bins. Star-forming
DRGs are plotted in blue points, quiescent DRGs are plotted in red
points. The blue lines represent the best-fit of the main sequence
of star-forming DRGs, and the blue dashed lines represent the
1$\sigma$ dispersion.
For comparison, the best-fit lines from Daddi et al. 2007 and Elbaz
et al. 2007 are also plotted.}
\end{figure*}

\subsection{Main sequence}%4.3

Recent studies have shown that there exists a correlation between star
formation rate and stellar mass of galaxies at different redshifts
($0<z<2$), and it is usually called ``main sequence" \citep{Brinchmann2004, Daddi2007, Elbaz2007, Pannella2009, Rodighiero2011, Barro2013, Speagle2014, Whitaker2014}. To investigate the
essential star formation status of DRGs, we show the ``main sequence"
relation in the four redshift bins in Figure 8. In this figure,
the blue and red points represent star-forming and quiescent DRGs,
classified by the rest-frame UVJ colors, respectively. To specify
the dependence of SFR on stellar mass, we also plot a linear-fit line
(blue solid line) and its $1\sigma$ dispersion (blue dashed line) for
star-forming DRGs in each redshift panel. For comparison, we overplot
the best-fit ``main sequence" relation for $z\sim2$ galaxies from
\cite{Daddi2007} and $z\sim1$ galaxies from \cite{Elbaz2007}. From this
figure, we find a positive correlation between SFR and $M_{*}$ of DRGs
in all four redshift bins. The slopes of the SFR$-M_{*}$ correlation of
star-forming DRGs in different redshift panels are similar. At fixed
stellar masses, the SFRs of galaxies on the SFR$-M_{*}$ correlation
increase with redshifts, which indicates that star-forming DRGs were
much more active on average in the past, due to greater abundance of gas
in early universe.

\begin{figure*}% Fig.9
\includegraphics[angle=-90,width=0.95\textwidth]{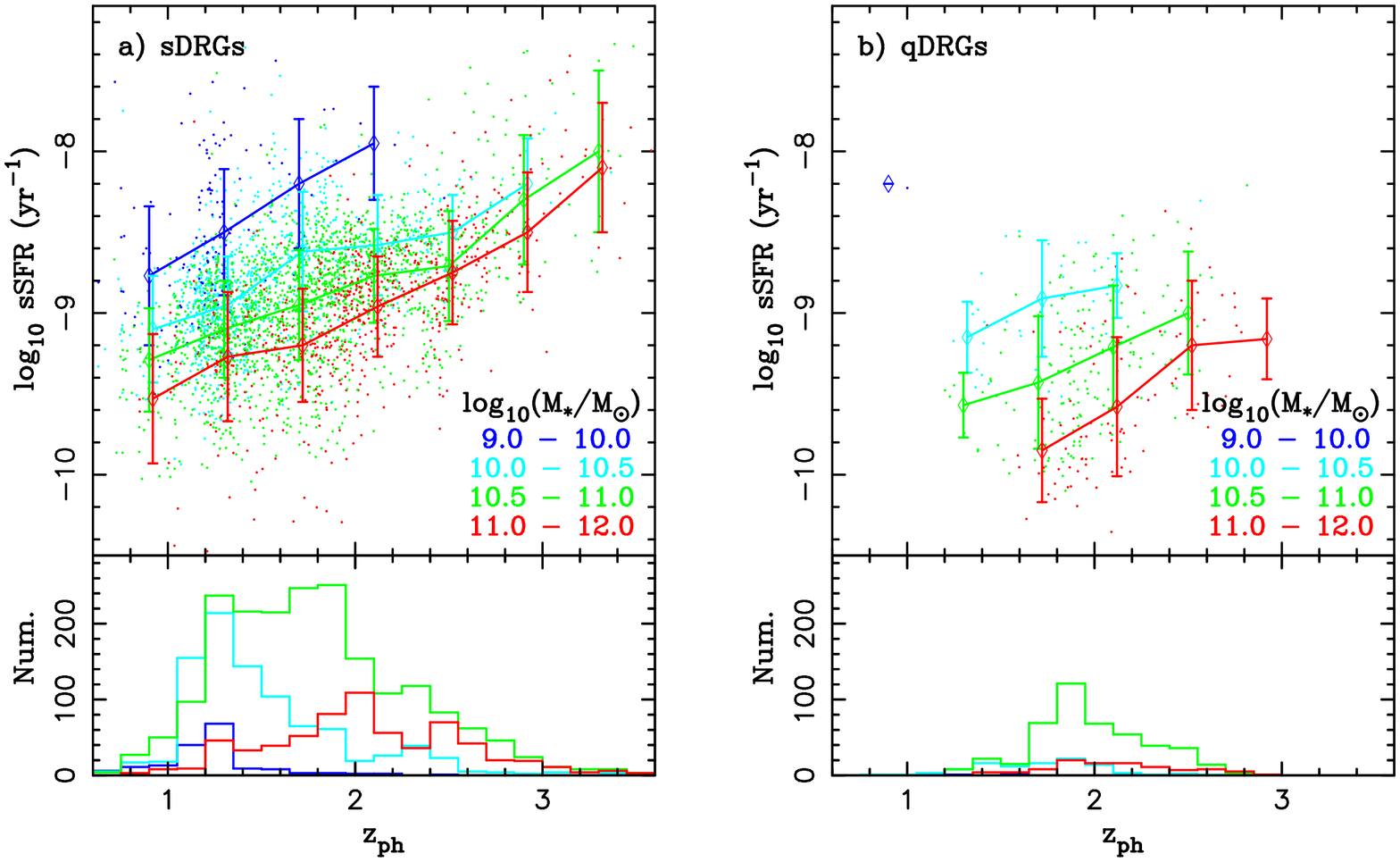}
\caption{sSFR vs. \zph\, diagram for star-forming DRGs in left
panel and quiescent DRGs in right panel. DRGs in different mass bins are
plotted with different colors.
The diamond symbols represent average sSFR value in redshift bin
$\Delta\zph=0.4$, and the error bars represent its
1$\sigma$ deviation. The histograms in the bottom panels show the redshift
 distributions of DRGs.}
\end{figure*}

Figure 9 shows the dependence of specific SFR (sSFR, $sSFR=SFR/M_{*}$)
on redshifts in four $M_{*}$ bins.
The $sSFR-\zph$ relation of star-forming and quiescent DRGs are shown
in the left and right panel, respectively.
In each panel, we divide DRGs into four $M_{*}$ bins ({\bf see Figure 9}).
For each $M_{*}$ bin, we calculate the $\langle$sSFR$\rangle$ (diamond
symbols) and its standard deviation (error bars) in each redshift bin
($\Delta\zph=0.4$). We find the sSFRs of DRGs increase with redshifts in all stellar mass bins, and DRGs with higher stellar masses generally have lower sSFRs. This is most likely due to the fact that the universe at
early time is more active, and massive galaxies formed most of their
stars earlier and slowed down more rapidly than their low-mass counterparts,
star formation contributes more to the growth of low-mass galaxies than
high-mass galaxies. This result is in agreement with the ``downsizing" mass-dependent quenching scenario \citep{Kajisawa2010,Kajisawa2011}.

\begin{figure*}% Fig.10
\includegraphics[angle=-90,width=0.95\textwidth]{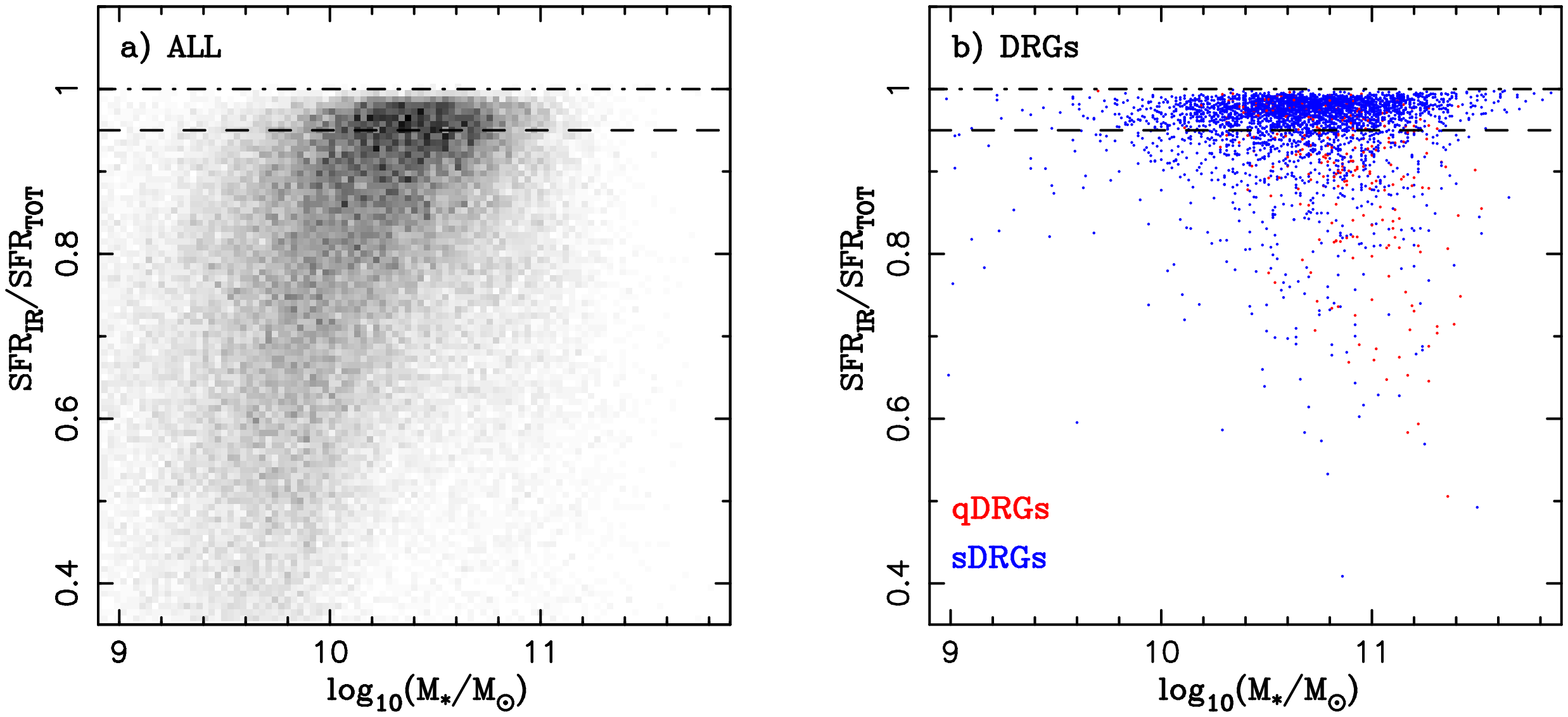}
\caption{SFR$_\mathrm{IR}$/SFR$_\mathrm{tot}$ vs. stellar mass ($M_{*}$)
diagram for DRGs (right panel) and all galaxies in the COSMOS/UltraVISTA
field (left panel). The dashed line corresponds to
SFR$_\mathrm{IR}$/SFR$_\mathrm{tot}=95\%$.}
\end{figure*}

\subsection{Star forming status and [3.6]--[24] color}%4.4

The total SFR of galaxy we used are determined via
SFR$_\mathrm{tot}$=SFR$_\mathrm{UV,uncorr}$+SFR$_\mathrm{IR}$
\citep{Muzzin2013a}. The SFR$_\mathrm{UV,uncorr}$ comes from SED fitting
method and the SFR$_\mathrm{IR}$ is derived by 24 $\mu$m flux. Higher
fraction of SFR$_\mathrm{IR}$ in SFR$_\mathrm{tot}$ means there are more
UV photons absorbed by dust, and these energies are re-emitted
from IR range. We plot the SFR$_\mathrm{IR}$/SFR$_\mathrm{tot}$
vs. $M_{*}$ of all galaxies (left panel) and DRGs (right panel, red
points for qDRGs, and blue points for sDRGs) in Figure 10.
From this figure we find there exists a
trend in the SFR$_\mathrm{IR}$/SFR$_\mathrm{tot}$ vs. $M_{*}$
diagram of all galaxies: high-mass galaxies tend to have higher
SFR$_\mathrm{IR}$/SFR$_\mathrm{tot}$ ratios, indicating that
massive galaxies tend to contain more dust and they are more metal
rich, thus the extinction for UV lights are heavier than low-mass
galaxies. Similar results are also found by \cite{Tadaki2013} in
their H$\alpha$ emitter (HAE) sample. We also find DRGs occupy the
high-mass and high ratio of SFR$_\mathrm{IR}$/SFR$_\mathrm{tot}$
region, which suggests the $J-K$ color criterion effectively selects
the massive dusty sources whose SFR$_\mathrm{IR}$ are dominant. As the
SFR$_\mathrm{IR}$/SFR$_\mathrm{tot}$ of most of DRGs great than 95\%,
it is suitable for DRGs using the 24 $\mu$m derived SFR$_\mathrm{IR}$
to trace the whole star formation activity.

There are 29.4\% quiescent and 49.8\% star-forming all $K$-selected galaxies in our sample have both [3.6] and [24] detection, and there are 37.3\% quiescent and 87.7\% star-forming DRGs have been detected at [3.6] and [24] bands simultaneously. The origins of the 24$\mu$m fluxes of the quiescent population can be due to processes unrelated to ongoing star formation, such as AGNs, cirrus dust heated by old stellar populations and circumstellar dust. \cite{Fumagalli2014} treat each of these components separately and compare their contributions to $L_{\rm IR}$ with the observed stacked values of $L_{\rm IR}$, and conclude that various processes other than star formation can contribute to the observed mid-IR flux, thus the 24$\mu$m-derived SFRs for quiescent galaxies should be the upper limits.

\begin{figure*}% Fig.11
\includegraphics[angle=-90,width=0.95\textwidth]{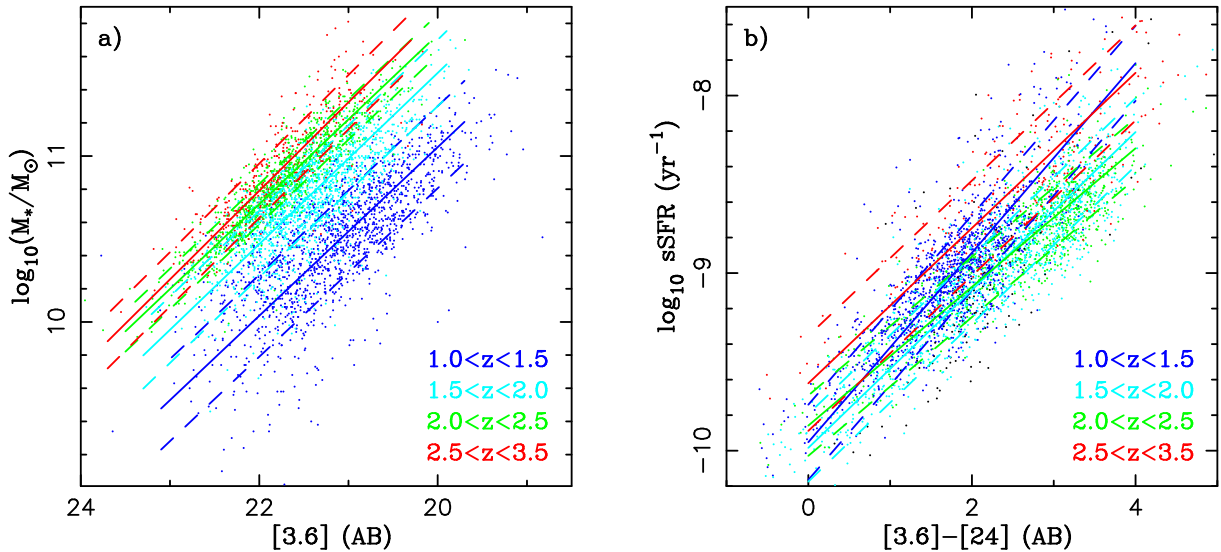}
\caption{Left panel: correlation between stellar mass and 3.6 $\mu$m
magnitude of DRGs. Right panel: correlation between sSFR and [3.6]--[24]
color of DRGs. DRGs in different redshift bins are plotted in different
colors. Solid lines are best-fit lines for each redshift bin, dashed
lines represent 1$\sigma$ deviation.}
\end{figure*}

At the redshift $z\sim1-3$, the IRAC 3.6 $\mu$m
probes the rest-frame NIR range, which can be derived into stellar
mass \citep{Bell2001, Cole2001}. The left panel of Figure 11 shows the
relationship of [3.6] vs. $M_{*}$ of all DRGs. We divide DRGs into four
redshift bins, and find there exists a strong correlation between [3.6]
and $M_{*}$. The linear-fit standard deviations of four redshift bins
(from low to high) are 0.25, 0.20, 0.12 and 0.16, respectively. Similarly,
the strong correlation between sSFRs and [3.6]--[24] colors in wide
redshift range ($1<z<3.5$) are also found in the right panel of Figure
11. The standard deviations of linear fit of four redshift bins (from
low to high) are 0.21, 0.19, 0.17 and 0.27, respectively. The average
[3.6]--[24] color of quiescent all $K$-selected galaxies is 0.78, while star-forming
 all $K$-selected galaxies is 2.03. For our DRG sample, the average [3.6]--[24] color
of quiescent DRG is 1.57, while star-forming DRG is 2.25. The [3.6]--[24] color of most of these 24$\mu$m-detected quiescent DRGs are generally consistent with but slightly redder than that of QGs presented by \cite{Brammer2009}, \cite{Wang2012} and \cite{Huang2013}, who identify QGs with the empirical [3.6]--[24] color criterion which is equivalent to a sSFR$\sim 10^{-10} \rm (yr^{-1})$. Quiescent and star-forming populations separate obviously in both DRG and all $K$-selected galaxy samples, which implies that [3.6]--[24] color is a good indicator for tracing star-forming status of galaxies. The similar results were also described in \cite{Huang2013}.

\begin{figure*}% Fig.12
\includegraphics[width=0.99\textwidth]{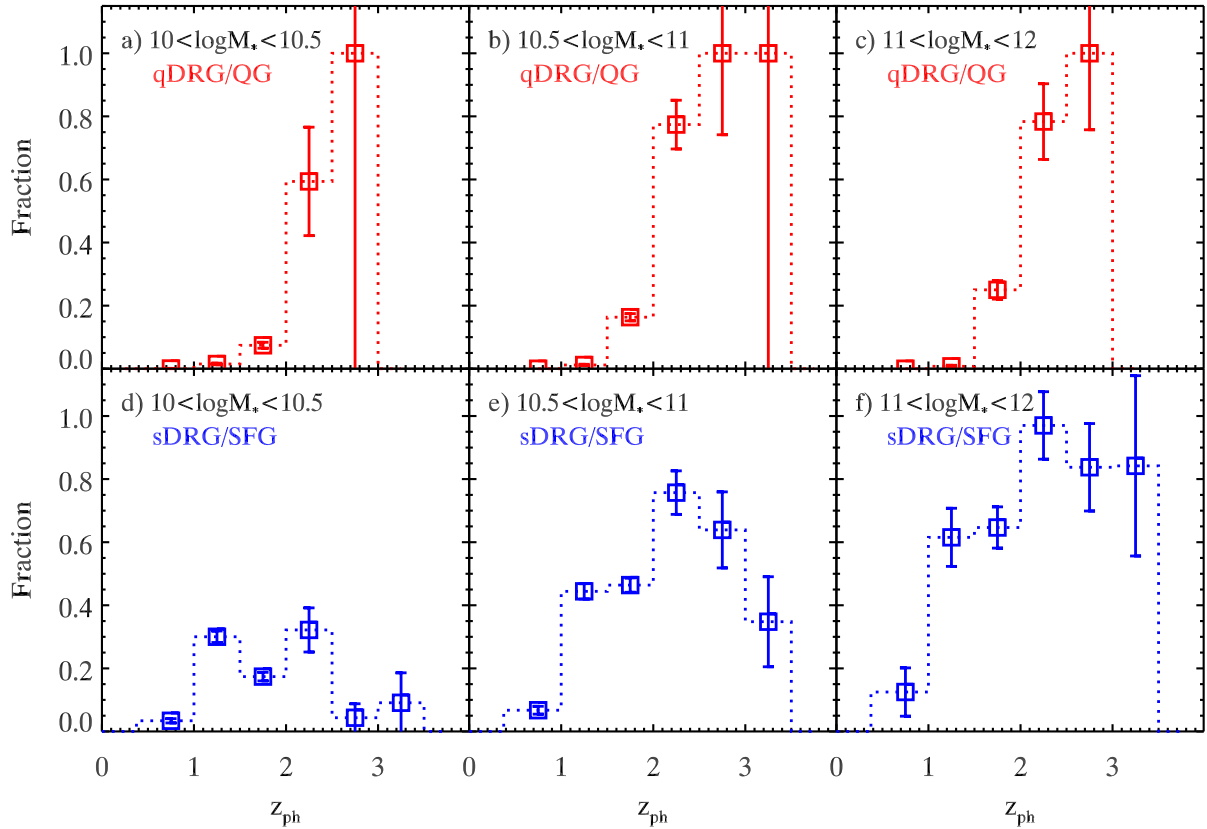}
\caption{Panel a), b) and c): Fraction of quiescent galaxies selected by the DRG criterion as a function of redshift. Panel d), e) and f): Fraction of dusty star-forming galaxies with $\rm SFR_{IR}/SFR_{tot}>0.95$ selected by the DRG criterion as a function of redshift. The fractions of QGs and SFGs are shown in three stellar mass bins ($\rm log$$_{10}(M_*/M_\odot)$=10-10.5, 10.5-11 and 11-12), and in each mass bin we divide them into six redshift bins ($z$=0.5-1, 1-1.5, 1.5-2, 2-2.5, 2.5-3 and 3-3.5) to calculate the fractions.}
\end{figure*}

To demonstrate how the DRG selection picks up each type of galaxies, we show the fraction of quiescent galaxies and the fraction of dusty star-forming galaxies with $\rm SFR_{IR}/SFR_{tot}>0.95$ selected by the DRG criterion as a function of redshift in Figure 12. Based on the stellar mass distribution of DRGs, we show the fractions for QGs and SFGs in three mass bins: $10<\rm log$$_{10}(M_*/M_\odot)<10.5$, $10.5<\rm log$$_{10}(M_*/M_\odot)<11$ and $11<\rm log$$_{10}(M_*/M_\odot)<12$, which are represented by the three columns of Figure 12 from left to right. The panels in the top and bottom row are fractions for QGs and SFGs, respectively. From this figure we find that (1) Higher fraction of dusty SFGs with $\rm SFR_{IR}/SFR_{tot}>0.95$ will be selected as DRG in higher mass bins, while the fractions of QGs remain nearly constant from low to high mass ranges; (2) Both fractions of QGs and SFGs increase with redshift in each stellar mass bin, but the increase for QGs are more dramatic: very few QGs can be selected as DRG at $z<2$, while at $z>2$ the fraction suddenly rises to a very high level. The result shown by this figure indicates that heavy dust reddening of $J-K$ color exists among high-mass SFGs, while the $J-K$ color of QGs is irrelevant to mass, because there is no dust extinction, the dramatic increase of the fraction of QGs is due to the sharp Balmer or 4000 \AA{} breaks of older stellar population shift into $J$-band at $z>2$.

\subsection{Rest frame U--V color vs. stellar mass}%4.5

\begin{figure*}% Fig.13
\includegraphics[angle=-90,width=0.95\textwidth]{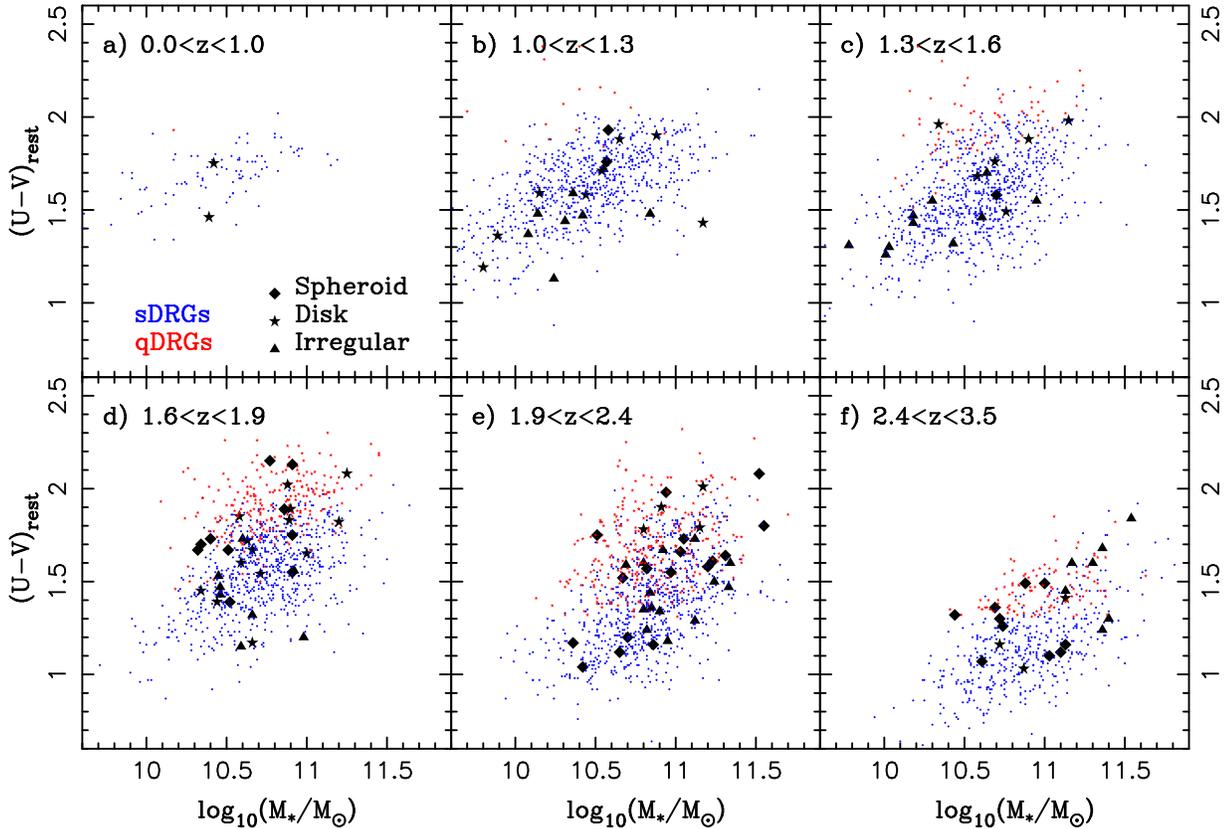}
\caption{Rest-frame $(U-V)$ color vs. stellar mass diagram for DRGs in
six redshift bins. Star-forming DRGs and quiescent DRGs are plotted in
blue and red points, respectively. DRGs which have visual morphological
types are also plotted in corresponding symbols.}
\end{figure*}

Galaxies present bi-modal distribution on the rest-frame color vs. mass
diagram (CMD), and they show red rest-frame colors due to dust extinction
or old stellar ages, thus the bi-modal distribution on CMD suggests
two distinct types of galaxy formation histories \citep{Bell2004,
Cassata2008, Brammer2009}. \cite{Talia2013} shows the good correlation
between $(U-B)_\mathrm{rest}$ and stellar mass, galaxies with redder
$(U-B)_\mathrm{rest}$ colors are more massive, no matter what kind of
morphologies they are. \cite{Wang2012} find that disk galaxies tend to be
more massive and have redder $(U-V)_\mathrm{rest}$ colors than irregular
galaxies in their IERO sample. \cite{DT2013} showed that with a particular
$(u-B)$ color, the high-redshift galaxies had on average higher sSFRs.

Figure 13 shows the distribution of DRGs on $(U-V)_\mathrm{rest}$
vs. $M_{*}$ diagram in six redshift bins. Star-forming and quiescent
DRGs are plotted in blue and red, respectively. DRGs having visual
morphological types introduced in Section 3.3 are also plotted in
corresponding symbols. We find that the $(U-V)_\mathrm{rest}$ color and $M_{*}$ of DRGs
correlate very well in all redshift bins, which is in good agreement with
Figure 10: more massive galaxies have redder $(U-V)_\mathrm{rest}$ colors due
to their larger internal dust extinctions or old stellar ages, which is also presented by \cite{KY2006}. However,
star-forming and quiescent DRGs cannot be separated efficiently on
CMD, they do not present clear bimodality, which is similar to what
\cite{Wang2012} and \cite{Huang2013} have found. DRGs with similar
stellar masses have bluer $(U-V)_\mathrm{rest}$ colors at high redshifts,
implying DRGs evolve with time, they gradually consume matter materials
and eventually slow down their star forming activities \citep{Kajisawa2005}. We also find that the irregular DRGs dominate the reddest $(U-V)_\mathrm{rest}$ color and high-mass end at high-redshifts, but gradually turn bluer and less massive than disk and spheroid DRGs toward low-redshifts. At $z>1.6$, there are many blue spheroid DRGs, they are in the stage of compact star-forming galaxies, which are supposed to be the progenitors of compact QGs at high redshift. The findings above may indicate a possible evolution scenario that massive disk-like DRGs merge with each other frequently by dissipative major merger at high redshifts, and produce massive dust-obscured irregular star-bursts, then through violent relaxation and angular momentum loss, they become spheroid-like DRGs and then quenched at later epoch. The findings are also supported by \cite{Bedregal2013}, \cite{Barro2013}, \cite{Bassett2013} and \cite{Williams2014}.

\section{Compare DRG to other color-selected high-redshift galaxies}%section5

\begin{figure}% Fig.14
\includegraphics[angle=-90,width=0.95\columnwidth]{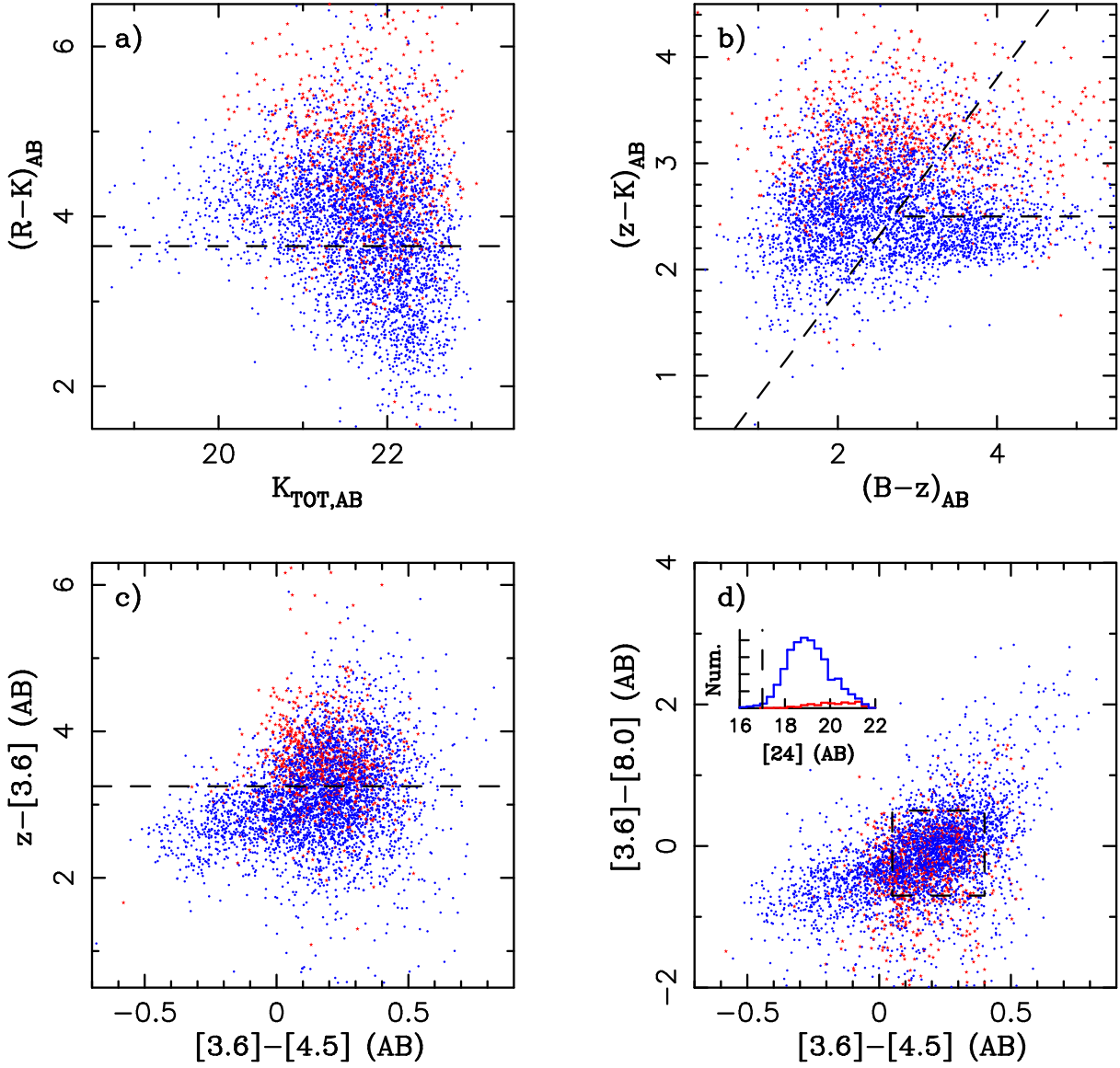}
\caption{(a): Distribution of star-forming DRGs (blue) and quiescent DRGs (red) on ERO selection diagram. The selection criteria are introduced in Simpson et al. 2006 and Lane et al. 2007. (b): Distribution of DRGs on BzK selection diagram. The selection criteria are defined by Daddi et al. 2004. (c): Distribution of DRGs on IERO selection diagram. The selection criteria are defined by Yan et al. 2004. (d): Distribution of DRGs on high-$z$ ULIRG selection diagram. This selection criteria are defined by Huang et al. 2009. The insert panel show the distribution of DRGs on the Spitzer 24 $\mu$m magnitude.}
\end{figure}

Over the past decade, many color selection techniques, which are designed
based on rest-frame spectral energy distribution features, are applied to
identify high-redshift galaxies \citep{Steidel1996, Smail1997,  Franx2003,
 Daddi2004, Gronwall2007,
Yan2007, Dey2008, Dunne2009, Huang2009,Caputi2012, Wang2012,
 Arcila2013, Ilbert2013}. However, due to the small sizes of
samples used in previous works, the overlaps between DRG and other color
selected population have not been well studied. In this section, we use
our large DRG sample to study the overlaps between DRGs and EROs, BzKs,
IEROs and high redshift ULIRGs.

First of all, we study the overlap between DRGs and EROs. To
be consistent with \cite{Kong2006} and \cite{Lane2007}, we use
$R-K$ color to select EROs from DRG sample. The color criterion
$(R_\mathrm{Subaru}-K_\mathrm{UKIDSS})_\mathrm{vega}>5.3$ in
\cite{Lane2007} can be converted to $(R-K)_\mathrm{AB}>3.65$
via $R_\mathrm{AB}=R_\mathrm{vega}+0.22$ and
$K_\mathrm{AB}=K_\mathrm{vega}+1.87$. Figure 14 (a) shows the distribution
of DRGs onto $(R-K)$ vs. $K$ diagram. We find that the distribution on
$K$ of quiescent DRGs is similar to star-forming DRGs. However, the
distributions of quiescent and star-forming DRGs  on $R-K$ color are
obviously different, quiescent DRGs have much redder $R-K$ colors than
star-forming DRGs, which attributes to the SED features of old stellar
populations in quiescent DRGs. 69\% DRGs can be selected as ERO,
which represents the low redshift range of DRG population. Many DRGs especially sDRGs distribute at higher redshift are missed by the ERO selection.

Secondly, we follow the procedure of \cite{McCracken2010} to perform $B-z$
color calibration and correct the $K$ band accounting for
$K_\mathrm{UltraVISTA}-K_\mathrm{WIRCam}=0.051$ described in
\cite{Muzzin2013a}, to see how DRGs satisfy the BzK criteria.
Figure 14 (b) shows
the distribution of DRGs on the $(z-K)$ vs. $(B-z)$ plane.
Star-forming DRGs
are slightly bluer than quiescent DRGs on $B-z$ color, but are obviously
bluer on $z-K$ color. 60.4\% DRGs could be selected as sBzKs,
17.7\% DRGs could be selected as pBzKs. Other DRGs mainly distribute in the
redshift range of $z<1.4$, most (73\%) of which are EROs.

Thirdly, we analyze the overlap between DRGs and IEROs color criterion
which is introduced by \cite{Yan2004}. The purpose of IERO method is
designed for selecting massive galaxies with mature stellar populations
at $z\sim2$ based on 4000 \AA\, Balmer break shifting beyond $z$-band or
steep slope of the rest-frame UV spectra of galaxies. Thus, a simple
color cut $z$--[3.6] is designed to select high redshift galaxies. Young
and blue galaxies cannot reach this threshold even at higher redshifts
\citep{Wang2012}. Based on the simulation performed by continuous star
formation (CSF) models in BC03 code,
we obtain $z_\mathrm{ACS}-z_\mathrm{Subaru}=-0.11$ to calibrate
$z$-band. As shown in Figure 14 (c), distributions on [3.6]--[4.5] color
of two population of DRGs are almost the same. However, star-forming
DRGs are obviously bluer than quiescent DRGs on $z$--[3.6] color. There
are 72\% quiescent DRGs and 39\% star-forming DRGs could be selected
as IERO. Quiescent DRGs are more likely to be selected as IERO, the
IERO color criterion rules out a large proportion of star-forming DRGs,
most of which are low-mass star-forming galaxies at low redshifts.

\begin{figure}% Fig.15
\includegraphics[angle=-90,width=0.95\columnwidth]{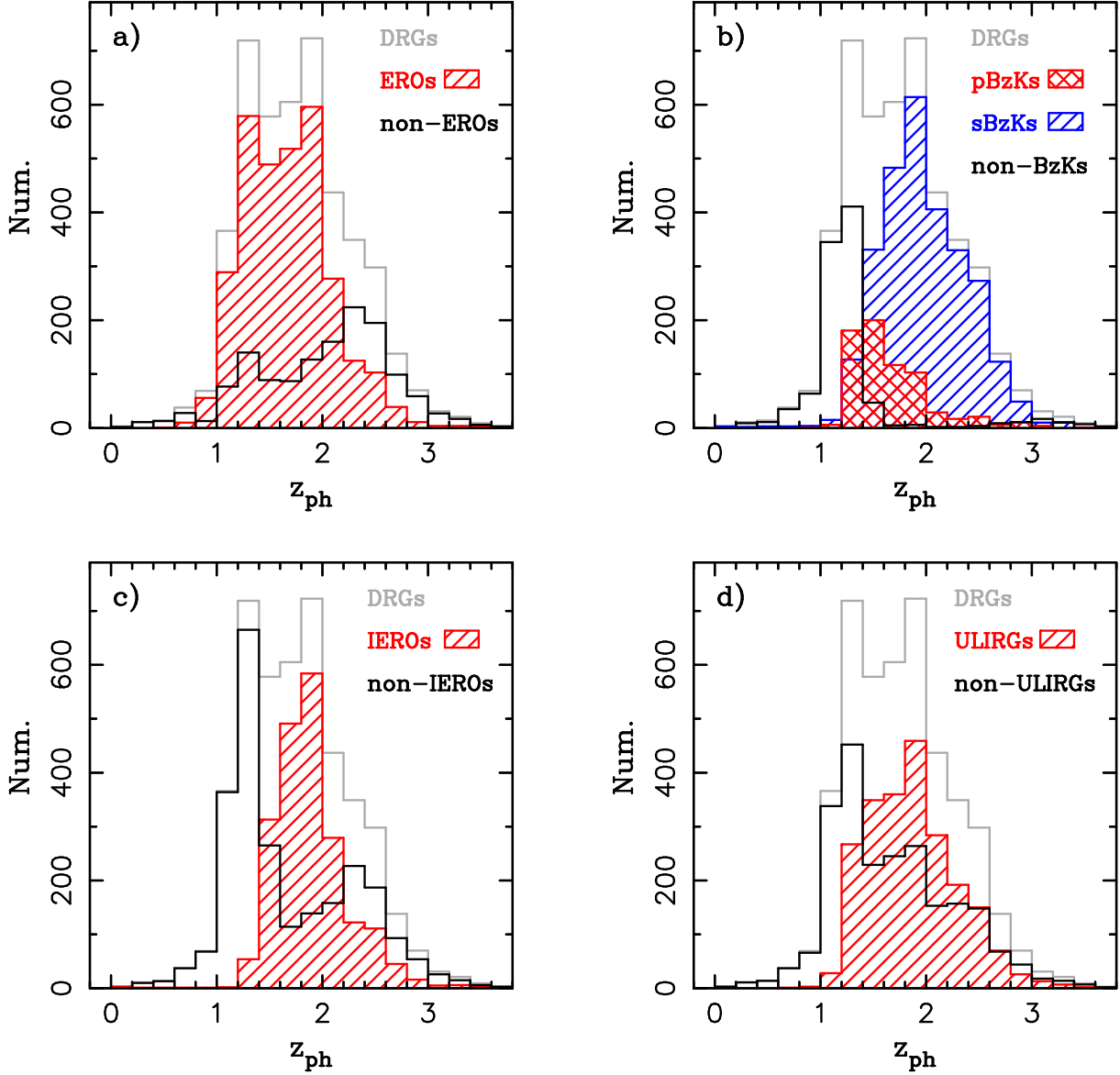}
\caption{Redshift distributions of DRGs that satisfy the ERO, BzK, IERO
and high-$z$ ULIRG color criteria. Red lines represent the distribution
of DRGs satisfying the corresponding color criterion in each panel and
black lines represent remanent DRGs.}
\end{figure}

Lastly, we perform comparison between DRG and high-$z$ ULIRG color
criteria introduced by \cite{Huang2009}. This IRAC color selection method
is designed for selecting massive, star-forming dominated ULIRGs at
$z\sim2$ based on the 1.6 $\mu$m stellar emission bump feature in galaxy
SEDs. In $1.5<z<3$, the four IRAC bands probe the rest-frame NIR bands
where galaxy SEDs have similar shape, thus the IRAC colors are robust in
determining redshift in this range \citep{Huang2009}. The color criterion
is set as $0.05<[3.6]-[4.5]<0.4$ and $-0.7<[3.6]-[8.0]<0.5$ based on the
M82 SED model \citep{Huang2004}. Figure 14 (d) shows the distribution of
DRGs on the [3.6]--[8.0] vs. [3.6]--[4.5] diagram. We find that the peaks of
DRGs on both two colors fall into the selection region designed by Huang et
al. 2009, 49.2\% DRGs can be selected by this rectangle region, which reflects this
selection method is efficient in determining redshifts. However, there are
only 1.3\% of DRGs satisfy their third cut, F$_\mathrm{24\mu m}>0.5$ mJy,
which is correspond to 17.15 mag in AB, as shown in the insert panel of
Figure 14 (d). Figure 15 shows the redshift distributions of DRGs, and the subsample of DRG
satisfying other color criteria. We find that a large proportion of DRGs could also
be selected as ERO, they mainly distribute at $z\sim1$, most of which
are dust reddened star-forming galaxies. The BzK method is very sharp
in determining redshifts. Most of DRGs in our sample can be selected
as sBzKs or pBzKs which distribute at $z>1.4$, non-BzKs mainly distribute
at $z<1.4$. The IERO criterion introduced by \cite{Wang2012} and the
high-$z$ ULIRG criterion introduced by \cite{Huang2009} are all efficient
in selecting $z\sim2$ galaxies.

\begin{figure}% Fig.16
\includegraphics[width=0.99\columnwidth]{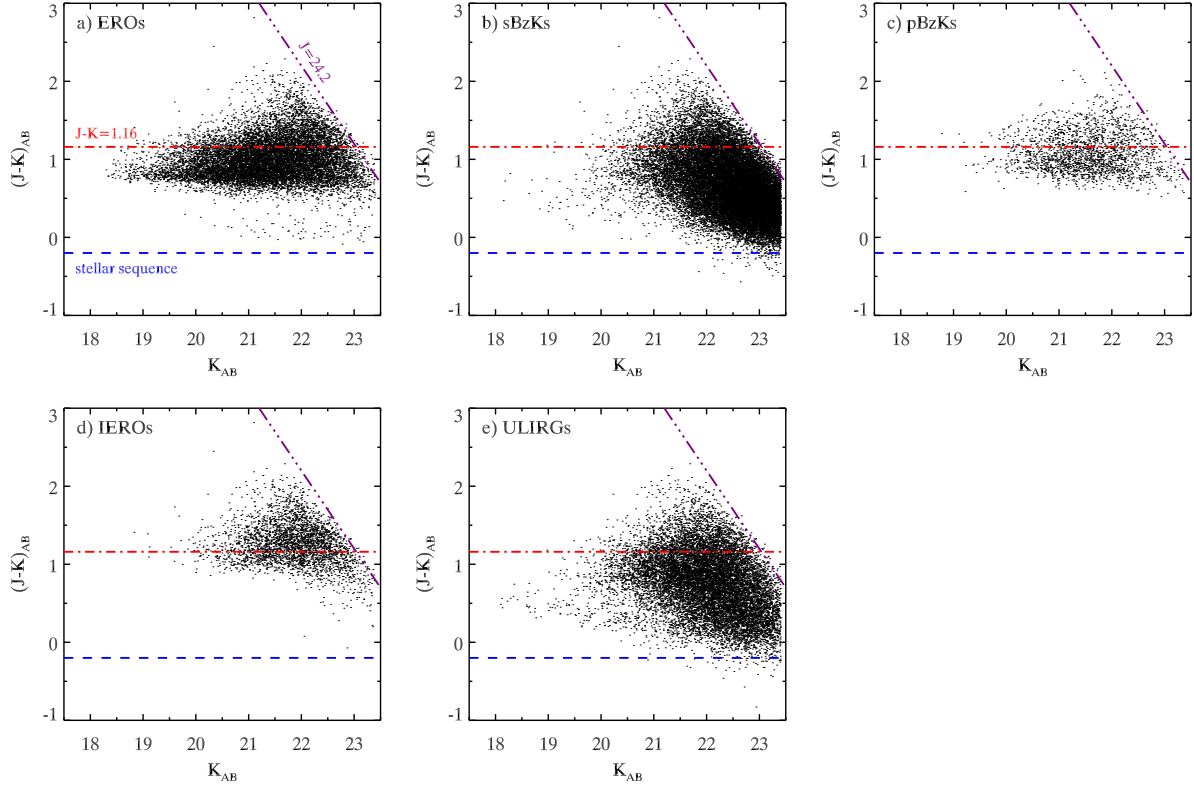}
\caption{Distributions of other color-selected galaxies on the DRG selection diagram. Panel a), b), c), d) and e) shows the distribution of ERO, sBzK, pBzK, IERO and high-$z$ ULIRG, respectively.}
\end{figure}

Similarly, we show how other color-selected galaxies satisfy the DRG selection criterion in Figure 16. Panel a), b), c), d) and e) shows the distribution on the DRG selection $J-K$ vs. $K$ diagram for ERO, sBzK, pBzK, IERO and ULIRG, respectively. 24.3\% EROs, 10.7\% sBzKs, 34.4\% pBzKs, 64.1\% IEROs and 15.8\% high-$z$ ULIRGs (not include the 24$\mu$m flux criterion) can be selected as DRG by the criterion $J-K>1.16$. From the comparisons performed above, we can get a clear point of view in different color selection criteria: methods designed for selecting galaxies at $z\sim2$ have many overlaps between each other. The DRG population is not a special galaxy group, the redshift distribution of them is broad, and most of them can also be selected by other color criterion.

\section{Summary}%section6

In this paper, we have described the construction of a large DRG sample
in the COSMOS/UltraVISTA field. Based on multi-band photometry data,
we analyze the physical properties of DRGs. Our main conclusions are
as follows.

1. We select a sample of 4485 DRGs with $(J-K)_\mathrm{AB}>1.16$
and $K_\mathrm{AB}<$23.4 mag, and identify 3725 star-forming
DRGs and 760 quiescent DRGs based on rest-frame UVJ colors. We
find the redshift distributions of DRGs are broad, ranging from
0.5 to 3.5 with a mean value of 1.79, but most of them (95.5\%)
distribute at $1<z<3$. The stellar masses of DRGs mainly distribute
at $10^{10}M_{\odot}-10^{11.5}M_{\odot}$. DRGs represent 70\% of all
galaxies in higher-mass bins at $z>2$, indicating that the DRG selection
criterion is sensitive to select massive galaxies at high redshift .

2. We study the morphological properties of DRGs in our sample, employing
data from $HST$ WFC3/F125W and F160W imaging within the COSMOS field. For
DRGs distribute at $1<z<1.6$ and $1.6<z<3$, we measure Gini and $M_{20}$
from F125W and F160W images, respectively. The morphological results
are consistent with our rest-frame UVJ color classification: quiescent
DRGs are generally compact and star-forming DRGs tend to have extended
structures, implying that these galaxies have experienced different
formation processes.

3. The SFR and $M_{*}$ of star-forming DRGs present good ``main sequence''
relations in all redshift bins. Moreover, the sSFR of DRGs increase
with redshift in all stellar mass bins and DRGs with higher stellar
masses generally have lower sSFRs, which indicates that galaxies were
much more active on average in the past, and star formation contributes
more to the mass growth of low-mass galaxies than to high-mass galaxies,
owning to massive galaxies formed most of their stars earlier and rapidly
slow down than their low-mass counterparts. This result is in agreement
with the ``downsizing" scenario.

4. There exists a trend in the SFR$_\mathrm{IR}$/SFR$_\mathrm{tot}$
vs. $M_{*}$ diagram of all galaxies: high-mass galaxies tend to have
higher SFR$_\mathrm{IR}$/SFR$_\mathrm{tot}$ ratios, indicating that
massive galaxies tend to contain more dust, thus the extinction for
UV lights are heavier than low-mass galaxies. We find DRGs occupy the
high-mass and high ratio of SFR$_\mathrm{IR}$/SFR$_\mathrm{tot}$ region,
which suggests the $J-K$ color criterion effectively selects the massive
dusty sources whose SFR$_\mathrm{IR}$ are dominant, and it is suitable
for DRGs using [3.6]--[24] color to trace their sSFRs. The fraction of QGs and the fraction of SFGs with $\rm SFR_{IR}/SFR_{tot}>0.95$ selected by DRG criterion as a function of redshift indicate that the red $J-K$ colors are mainly caused by dust reddening for SFGs and the Balmer or 4000 \AA{} breaks shifting into the $J$-band at $z>2$ for QGs.

5. The $(U-V)_\mathrm{rest}$ color strongly correlates $M_{*}$ of DRGs
in all redshift bins. Massive DRGs have redder $(U-V)_\mathrm{rest}$
colors and the $(U-V)_\mathrm{rest}$ colors of DRGs become bluer at
higher redshifts, suggesting high-mass galaxies have larger internal dust
extinctions or older stellar ages and they evolve with time. We also find that the irregular DRGs dominate the reddest $(U-V)_\mathrm{rest}$ color and high-mass end at high-redshifts, but gradually turn bluer and less massive than disk and spheroid DRGs toward low-redshifts. At $z>1.6$, there are many blue spheroid DRGs. They are in the stage of compact star-forming galaxies which are supposed to be the progenitors of compact QGs at high redshift. These findings may indicate a possible evolution scenario that massive
disk-like DRGs merge with each other frequently, then through violent
relaxation and angular momentum loss, they finally become spheroid-like
galaxies.

6. We study the overlaps between DRGs and EROs, BzKs, IEROs, high-$z$
ULIRGs, find DRGs in our sample can be selected by other color criteria
also. 69\% DRGs can be selected as EROs, and most of them have low
redshift. 60\% DRGs can be selected as sBzKs, 18\% DRGs can be selected as pBzKs,
non-BzKs mainly distribute at $z<1.4$, most (73\%) of them are EROs.
72\% quiescent DRGs and 39\% star-forming DRGs could be selected
as IERO, quiescent DRGs are more likely to be selected as IERO, the IERO
color criterion rules out a large proportion of star-forming DRGs, most of
 them are low-mass star-forming galaxies at low redshifts. 49\% DRGs can be selected by the rectangle region of high-$z$
ULIRG candidates and only  1.3\% of DRGs in our sample satisfy the
F$_\mathrm{24\mu m}>0.5$mJy flux cut, which introduced by \cite{Huang2009}. In turn, 24.3\% EROs, 10.7\% sBzKs, 34.4\% pBzKs, 64.1\% IEROs and 15.8\% high-$z$ ULIRGs (not include the 24$\mu$m flux criterion) can be selected as DRG by the criterion $J-K>1.16$.

\section*{Acknowledgments}

We are grateful to R. Abraham for access to his morphology analysis
code. We also acknowledge E. Daddi and A. Muzzin for their valuable comments. This work is based on observations taken by the CANDELS Multi-Cycle Treasury Program with
the NASA/ESA HST, which is operated by the Association of Universities for
Research in Astronomy, Inc., under the NASA contract NAS5-26555.
This work is supported by the National Natural Science Foundation of China (NSFC, Nos. 11303002, 11225315, 1320101002, 11433005, and 11421303), the Specialized Research Fund for the Doctoral Program of Higher Education (SRFDP, No. 20123402110037), the Strategic Priority Research Program ``The Emergence of Cosmological Structures" of the Chinese Academy of Sciences (No. XDB09000000), the Chinese National 973 Fundamental Science Programs (973 program) (2015CB857004), the Yunnan Applied Basic Research Projects (2014FB155) and the Open Research Program of Key Laboratory for Research in Galaxies and Cosmology, CAS.

%%references

\end{document}